\documentclass[runningheads]{llncs}
\usepackage{graphicx,xcolor}
\usepackage{bm}
\usepackage{amsmath}
\usepackage{enumitem}
\usepackage{ulem}

\usepackage[numbers,sort&compress]{natbib}

\usepackage{lineno}
\begin{document}
\nolinenumbers
\title{SMIXS: Novel efficient algorithm for non-parametric mixture regression-based clustering}

%
%
\author{Peter Mlakar \inst{1} \and
Tapio Nummi \inst{4} \and
Polona Oblak \inst{2}\and
Jana Faganeli Pucer\inst{3}}
\authorrunning{P. Mlakar et al.}
%
\institute{University of Ljubljana, Faculty of Computer and Information Science,\newline Republic of Slovenia, Slovenian Environment Agency \newline \email{peter.mlakar@gov.si} \and
University of Ljubljana, Faculty of Computer and Information Science
\email{polona.oblak@fri.uni-lj.si} \and
University of Ljubljana, Faculty of Computer and Information Science
\email{jana.faganeli@fri.uni-lj.si} \and
Tampere University, Faculty of Information Technology and \newline Communication Sciences \newline
\email{tapio.nummi@tuni.fi}}
\maketitle              
\begin{abstract}
We investigate a novel non-parametric regression-based clustering algorithm for longitudinal data analysis.
Combining natural cubic splines with Gaussian mixture models (GMM), the algorithm can produce smooth cluster means that describe the underlying data well.
However, there are some shortcomings in the algorithm: high computational complexity in the parameter estimation procedure and a numerically unstable variance estimator.
Therefore, to further increase the usability of the method, we incorporated approaches to reduce its computational complexity, we developed a new, more stable variance estimator, and we developed a new smoothing parameter estimation procedure.
We show that the developed algorithm, SMIXS, performs better than GMM on a synthetic dataset in terms of clustering and regression performance.
We demonstrate the impact of the computational speed-ups, which we formally prove in the new framework.
Finally, we perform a case study by using SMIXS to cluster vertical atmospheric measurements to determine different weather regimes.

\keywords{mixture models \and regression \and clustering \and smoothing splines}
\end{abstract}

\section{Introduction}

Longitudinal datasets contain samples described by measurements of one or more dependent variables over one independent variable, frequently denoted as time.
We collect such datasets with the intent of studying the time-dependent developmental nature of individual samples through the fluctuations in their measurements.
To this end, we can use regression techniques to provide insights regarding the dependence structure. 
Furthermore, we are able to extract additional information if subsets of samples share or exhibit similar developmental trends.
For such cases, mixture regression methods or clustering algorithms can prove useful.
Both of the aforementioned properties of longitudinal datasets are crucial for understanding the nature of the data, therefore the methods used to analyze these properties share similar importance.

We can conduct longitudinal data analysis in different frameworks, each pertaining to a different view of the problem.
One can approach the issue through the lens of generalized linear mixed models \cite{bolker2015linear,clskom}, k-means variants extended to perform better in a longitudinal setting \cite{clskml}, Bayesian methods \cite{clslu}, generalized estimation equations \cite{wang2014generalized}.
For comparison and overview of different longitudinal analysis techniques, we refer to \cite{doi:10.1080/03610918.2020.1861464,fitzmaurice2012applied,wu2006nonparametric}.
Here, we focus on a finite mixture model approach \cite{mclachlan2019finite,everitt2013finite}, due to its flexibility and statistical interpretability.

Specifically, we study the non-parametric regression-based clustering algorithm developed by \citet{nummi2018semiparametric}.
This algorithm leverages Gaussian mixture models (GMM) and smoothing splines to construct $c$ mixtures, described by smooth mean curves.
The smoothing splines constrain the individual mixture mean curves based on their estimated roughness.
We do this to control the effect noise would have on the final regressor since many real-world datasets contain noise in their measurements.
Additionally, the development in latent clusters is complicated and a general smooth function will provide a very good approximation of it.
We can control the amount of smoothing enforced per mixture with a smoothing parameter, rendering the algorithm flexible even when processing non-homogeneous data since each mixture is capable of adapting to specific parts of the dataset.
Therefore, the result of this procedure is the formulation of mixture means as smooth functions that continuously model transitions between measurements in an energy-optimal way \cite{green1993nonparametric}.
This differentiates the proposed algorithm from GMM, making it more resilient to strong noise signals present in the dataset, resulting in better clustering as well as regression performance compared to GMM.

However, some drawbacks mitigate the practical applicability of the algorithm \cite{nummi2018semiparametric}.
First, the exact estimation of this algorithm is subject to high computational complexity.
Because of the iterative nature of the optimisation procedure and the computation of matrix inverses which are required for the estimation of parameters, the algorithm quickly becomes intractable.
Adding to the computational burden is the smoothing parameter selection procedure, which requires multiple estimations in each optimization iteration.
Unrelated to the computational complexity woes, the variance parameter estimator exhibits unwanted numerical behaviour in the context of Expectation maximization (EM) \cite{mclachlan2019finite} and lastly, we were not able to find any comparative studies of the algorithm \cite{nummi2018semiparametric} with other existing methods which would further bolster the algorithm's usability.
Therefore, to overcome these issues, we propose a new algorithm SMIXS with the following major contributions:
\begin{itemize}[label=$\bullet$]
    \item Implementation of speed-ups for crucial computational bottlenecks.
    \item Alternative smoothing parameter selection procedure using gradient descent.
    \item Derivation of a more stable penalty-corrected variance estimator.
    \item Comparison of SMIXS against the base GMM in terms of regression performance, clustering performance, and computational complexity.
    \item An open source implementation of the algorithm in Julia and Python available on GitHub \cite{smixsgithub}.
\end{itemize}
The remainder of the paper is structured in the following manner.
We present the algorithm SMIXS in Section \ref{ch:mth}. 
In Section \ref{ch:rslt} we describe the conducted empirical analysis on a synthetic dataset, the results of which are provided in Subsection \ref{sec:syneval}.
We present the results of a case study in Subsection~\ref{ch:cstd} where we use SMIXS to cluster the atmospheric sounding data of temperature measurements. 
Additional derivations and more detailed explanations are supplemented in Appendix \ref{ch:appendix}.

\section{Mixture regression for longitudinal clustering}
\label{ch:mth}
   
The SMIXS algorithm builds upon the Gaussian mixture model, constraining its mixture means by using natural cubic smoothing splines.
This enables the modeling of a rich set of curves \cite{nummi2018semiparametric} while also better describing the underlying development in latent groups.
The parameters of the SMIXS model are estimated using the Expectation Maximization (EM) algorithm \cite{dempster1977maximum,mclachlan2019finite}.
Let us first define the notation used in the following sections:
\begin{itemize}[label=$\bullet$]
    \item Let $n$ denote the number of total samples and let vector $\bm y_i$ denote the $i$-th sample with $p$ elements. 
    \item Let $\Omega$ be the set of all model parameters and let $\omega_k$ denote the parameter subset belonging to the $k$-th mixture of the model, and let $c$ be the number of mixtures.  
    \item Each $\omega_k$ contains the following parameters: the mixture mean vector $\bm \mu_k$ with $p$ elements, mixture standard deviation $\sigma_k$, and mixing proportion $\pi_k$.
    The mixing proportions denote the relative importance of individual mixtures and their sum is equal to $\sum_{k = 1}^{c}\pi_k = 1$.
    \item To convert the Gaussian mixture fitting problem from an incomplete to a complete framework we use the random variable $z_{ik}$ to denote whether or not the $i$-th sample belongs to the $k$-th cluster. 
    If we knew the true values of these random variables for each individual, the clustering would be rendered trivial.
    Since we do not, we can look at this as a type of missing data problem which facilitates the estimation of the model parameters.
    The estimator of the variable $z_{ik}$, denoted as $\hat{z}_{ik}$ (hat symbol over a variable denotes its estimator), is computed as $\hat{z}_{ik} = E_{\Omega}(z_{ik}|\bm y_i) = \frac{\pi_k f_k(\bm y_i, \omega_k)}{\sum_{l = 1}^{c}\pi_l f_l(\bm y_i, \omega_l)}.$ Each $\hat{z}_{ik}$ is a positive real number.
    \item Let $f_k$ denote the $k$-th mixture probability density function which in our cases amounts to the multivariate Gaussian distribution with a diagonal covariance matrix $\sigma_k^2\bm I$.
    Therefore, within the mixture, we assume independence between measurements and homogeneous variance.
    \item Let the roughness matrix $\bm G$ be defined as $\bm G = \bm Q \bm R^{-1} \bm Q^\top$, where $\bf Q$ and $\bf R$ represent two band matrices.
    They encode the relationship between a smoothing spline's values with its second derivatives at the spline knots.
    For more details see \citet{green1993nonparametric}.
\end{itemize}
The quantity maximized during the EM maximization step, with respect to the SMIXS model parameters, is the penalized conditional expectation of the log-likelihood.
To define the penalty term we begin with the smoothing parameter $\lambda_k$, which controls the penalty's relative importance compared to the regression enforced by the conditional expectation.
The penalty term is then defined as
\begin{align}
\label{eq:pnlt}
    P_k = \lambda_k\bm\mu_k^\top\bm G\bm\mu_k,
\end{align}
and penalized conditional expectation is written as
\begin{align}
\label{eq:etilde}
    \tilde{E}(\Omega, \bm y) = \sum_{i = 1}^{n}\sum_{k = 1}^{c}\hat{z}_{ik}(\log (\pi_k) + \log (f_k(\bm y_i, \omega_k))) - \sum_{k = 1}^{c}P_k.
\end{align}
The penalty term $P_k$ restrains the individual mixture means based on their roughness.
We defined the concept of roughness as the definite integral of the squared second derivative of a smooth function interpolating the individual mixture mean elements.
The nature of this roughness penalty forces the mixture mean to take on the values of a natural cubic spline at the knots \cite{green1993nonparametric}.
\\
Continuing with the maximization step of EM, we are required to estimate the remaining model parameters.
We compute the mixture proportion estimators as the average of the $n$ index estimators for the $k$-th mixture
\begin{align*}
\hat{\pi}_k = \frac{1}{n}\sum_{i = 1}^{n} \hat{z}_{ik}
\end{align*}
and the cluster mean estimators can be calculated by
\begin{equation}
\label{eq:mu_est}
\hat{\bm \mu}_k = \left (\sum_{i = 1}^{n}\hat{z}_{ik}\bm I + \alpha_k\bm G \right )^{-1}\sum_{i = 1}^{n}\hat{z}_{ik}\bm y_i,
\end{equation}
where $\alpha_k = \frac{\lambda_k}{\sigma_k^2}$.
This smoothing weight substitution enables the computation of the mixture mean estimator without the direct need to calculate the variance.
After we determine $\alpha_k$ we can proceed with the estimation of other parameters, without any loss of generality, in the order of $\hat{\bm \mu}_k$, $\hat{\sigma}^2_k$.
For more details concerning the substitution refer to \citet{green1993nonparametric}.

The procedures with which we select the smoothing parameter $\alpha$ and compute the variance estimators differ from the ones proposed by \citet{nummi2018semiparametric}.
We introduce and provide arguments for the use of our approaches in the following sections.

\subsection{Variance estimator}

To calculate the variance estimator as defined in \citet{nummi2018semiparametric} we first select a smoothing parameter value, compute the corresponding mixture mean, and then estimate the variance of a multivariate Gaussian distribution, disregarding the penalty term's direct influence on the variance.
The rationale behind this is that when we estimate the mean of a mixture we considered the smoothing constraint.
Therefore, when we compute the variance, for which we require the mean, no additional consideration towards the smoothing parameter is needed as we applied it at the point of mean estimation.
However, we found that this yields unstable performance in certain cases.
To be exact, the expectation maximized during EM decreases in value over consecutive iterations.
This is unwanted behaviour in the EM framework \cite{wu1983convergence} and is due to the fact that the variance computed this way is not the maximum expected log-likelihood estimator.
To this end, we introduce the penalty-corrected variance estimator which we calculate by computing the maximum expected log-likelihood estimator for $\sigma_k^2$ using Equation (\ref{eq:etilde}).
This estimator is defined as
\begin{align*}
    \hat{\sigma}_k^2 = \frac{\sum_{i = 1}^{n}\hat{z}_{ik}(\bm y_i - \bm \mu_k)^{\top}(\bm y_i - \bm \mu_k) + \alpha_k\bm \mu_k^{\top}\bm G\bm \mu_k}{\sum_{i = 1}^{n} \hat{z}_{ik}p}.
\end{align*}
We provide a more detailed derivation in Appendix \ref{ch:appendix}.
The addition of $\alpha_k\bm \mu_k^{\top}\bm G\bm \mu_k$ to the numerator compensates for the smoothness constraint and mitigates the unwanted convergence behaviour.

\subsection{Alpha parameter selection}

There are multiple ways one can select the value of the smoothing parameter $\alpha_k$.
As a possible alternative \citet{nummi2018semiparametric} proposed the cluster-wise maximization of the so-called profile log-likelihood function with respect to the corresponding smoothing parameter $\alpha_k$.
This entails that the optimal smoothing parameter $\alpha_k$ is the one that minimizes the variance $\hat{\sigma}_k^2$.
However, we believe that such an estimator biases small $\alpha_k$ values as the variance is smallest when the mixture mean follows the weighted arithmetic average, entailing that the smoothing parameter equals zero.

Therefore, we propose the use of cross-validation which is also supported in the literature \cite{green1993nonparametric}.
Let $\hat{\bm \mu}_j^{-\{ij\}}=\hat{\bm \mu}_j^{-\{ij\}}(\alpha_k)$ denote the $k$-th mixture mean estimator, defined in Equation \eqref{eq:mu_est}, computed by omitting the $j$-th element of the sample $i$.
Then cross-validation can be written as
\begin{equation}
\label{eq:cv}
    CV(\alpha_k) = \sum_{i = 1}^{n}\hat{z}_{ik}\sum_{j = 1}^{p}(\hat{\bm \mu}_j^{-\{ij\}} -  \bm y_{ij})^2.
\end{equation}
This however requires a grid search over multiple $\alpha_k$, significantly decreasing the speed of the estimation procedure.
It is also not sensible to do a comprehensive grid search in the starting iterations of the EM algorithm since the remaining parameter initializations are not optimized.

To this end, we suggest that a gradient descent approach might be in order since it eliminates the need for evaluating many alpha values at each iteration of the EM algorithm.
By limiting $\alpha_k$ to the interval $[1, 10^6]$ and starting at $\alpha_k = 1$ we conduct gradient descent on the cross-validation score, doing one step each EM iteration. 
By approximating the derivative of $CV$ by its differential quotient and by denoting the update rate with $\theta$, we define the update of $\alpha_k$ to be
\begin{align*}
    \alpha_k^{new} = \alpha_k - \theta \frac{CV(\alpha_k + h) - CV(\alpha_k)}{h}.
\end{align*}
In our case we chose $\theta = 10^{-3}$ and $h = 0.1$.
This eliminates the grid search computational dilemma to a certain extent and at the same time allows us to reach satisfactory smoothness after multiple iterations.
We can also utilize a dynamic learning rate in this procedure, which might result in faster convergence.
However, one must note that complex learning rate estimators usually require additional non-trivial computations which might nullify the expected benefits.
For an example of a dynamic learning rate used in this context refer to \cite{mlakar2021application}.

\subsection{Computational complexity reduction}

There are two potentially problematic computational bottlenecks in the presented algorithm.
To alleviate these burdens, we identified appropriate solutions which we describe in this section.
For more details on both approaches refer to \citet{green1993nonparametric} and \citet{reinsch1967smoothing}, and to the Appendix \ref{ch:appendix}.

The first computational problem is computing the matrix inverse 
\begin{equation}\label{eq:S}
    \bm S = \left ( \sum_{i = 1}^{n}\hat{z}_{ik}\bm I + \alpha_k \bm G \right )^{-1}.
\end{equation}
in Equation \eqref{eq:mu_est}, which we call the smoothing matrix. 
We efficiently compute the inverse by using the Reinsch algorithm \cite{reinsch1967smoothing}. This enables us to express the matrix $\left (\sum_{i = 1}^{n}\hat{z}_{ik}\bm I + \alpha_k\bm G \right )$ in a more suitable form.
Define first the matrix $\bm W_k = \sum_{i = 1}^{n}\hat{z}_{ik}\bm I$ and the vector $\tilde{\bm y}_k = \sum_{i = 1}^{n}\hat{z}_{ik}\bm y_i$.
Also let the vector $\bm \gamma_k$ denote the vector of second order derivative of the natural cubic spline described by $\bm \mu_k$, evaluated at its knots.
This entails that we can implicitly express the mixture mean estimator in the following way
\begin{equation}\label{eq:R+QWQ}
    \bm Q^\top \bm W^{-1}_k \tilde{\bm y}_k = \left (\bm R + \alpha_k \bm Q^\top \bm W^{-1}_k \bm Q \right) \hat{\bm \gamma}_k.
\end{equation}

 Note that since $\bm R + \alpha_k \bm Q^\top \bm W^{-1}_k \bm Q $ is a symmetric, pentadiagonal, positive-definite matrix, we can efficiently compute its inverse by the use of its Cholesky decomposition. 
 This enables us to estimate $\hat{\bm \gamma}_k$ from Equation \eqref{eq:R+QWQ} and so we can compute the mixture mean estimator $\hat{\bm  \mu}_k$ from~\eqref{eq:mu_est} in linear time with respect to the number of measurements. 

To compute the cross-validation score the individual mixture mean vector would have to be estimated for each omission of one measurement from each sample. This renders the computation of the cross-validation score already cumbersome but one must not forget that this procedure is performed for each value of $\alpha_k$ we wish to evaluate.
To tackle this second problem we speed up the procedure by using
 \citet{hutchinson1985smoothing} algorithm. 
The difference $\hat{\bm \mu}_j^{-\{ij\}} -  \bm y_{ij}$ required for the estimation of the cross validation score can then be expressed as
\begin{equation}
\label{eq:diff}
    \hat{\bm \mu}_j^{-\{ij\}} -  \bm y_{ij} = \frac{\hat{\bm \mu}_j - \bm y_{ij}}{1 - \bm S_{jj}\hat{z}_{ik}},
\end{equation}
where $\bm S_{jj}$ are the diagonal elements of the smoothing matrix $ \bm S$.
For more details refer to Appendix \ref{ch:appendix}.
Note that Equation (\ref{eq:diff}) is a powerful statement as it lets us express the difference $\hat{\bm \mu}_j^{-\{ij\}} -  \bm y_{ij}$ in terms of the mixture mean estimator $\hat{\bm \mu}_k$, computed by not omitting any data from the dataset.
This entails that we require only one mixture mean estimation per $\alpha_k$ to calculate its corresponding cross-validation score for each omission.

\section{Empirical evaluation}
\label{ch:rslt}

We empirically evaluate and compare SMIXS to the base GMM algorithm, from which SMIXS is derived, on several synthetic datasets.
We evaluate its clustering performance, regression performance, and computational complexity. 
We implement our version of the GMM algorithm, which differs from the SMIXS implementation only in the maximization step. 
This way we make sure that the difference in algorithms is only due to the way SMIXS estimates the model parameters in the maximization step.
By comparing the performance of the two algorithms we want to show the advantages of SMIXS in terms of clustering and regression curve accuracy. 

The running times of both SMIXS and GMM algorithms depend on the number of clusters we wish to find, the number of subjects present in a dataset, and the number of measurements the dataset contains for each subject. 
The synthetic datasets enable us to investigate their performance by varying the input variables mentioned above.
Also, instead of using a static learning rate to conduct the smoothing parameter estimation in the synthetic dataset study, we use a dynamic one based on the approximation of the second derivative \cite{mlakar2021application}.
This might speed up the convergence of the estimation procedure since the learning rate is adjusted based on the slope of the gradient, albeit by spending additional resources by estimating the second derivative.

To show the applicability of the SMIXS algorithm to a real-world problem we cluster atmospheric sounding data from Ljubljana. 
Temperature inversions in Ljubljana are common in winter which greatly affects air quality. 
We hypothesize that a mixture analysis could improve the prediction of PM$_{10}$ concentrations.
In this case study, we use a static learning rate in the smoothing parameter estimation procedure, since it provides good clustering and regression performance.

\subsection{Algorithm initialization}
The EM algorithm is an iterative approach, whose performance is highly dependent upon the initial values of the involved parameters. 
The result of the EM algorithm is usually a local instead of the global maximum. 
The goal of finding good results necessitates the execution of multiple runs with different starting points. 
We evaluate the quality of each run using log-likelihood. 
To initialize the starting parameters of cluster means, cluster variances, and mixture proportions we utilize the k-means algorithm.

\subsection{Synthetic dataset construction}\label{sec:synth_data}

\begin{figure}[htb]
\begin{center}
\includegraphics[width=0.32\textwidth]{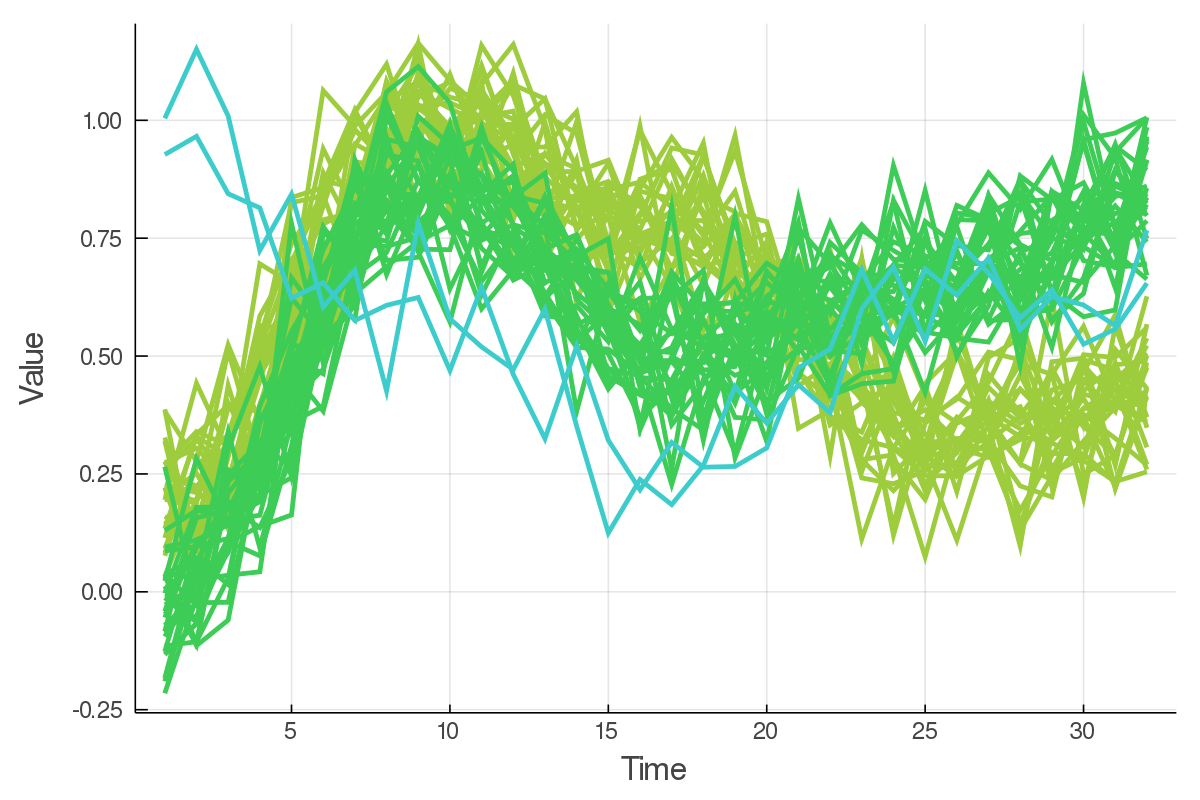}
\includegraphics[width=0.32\textwidth]{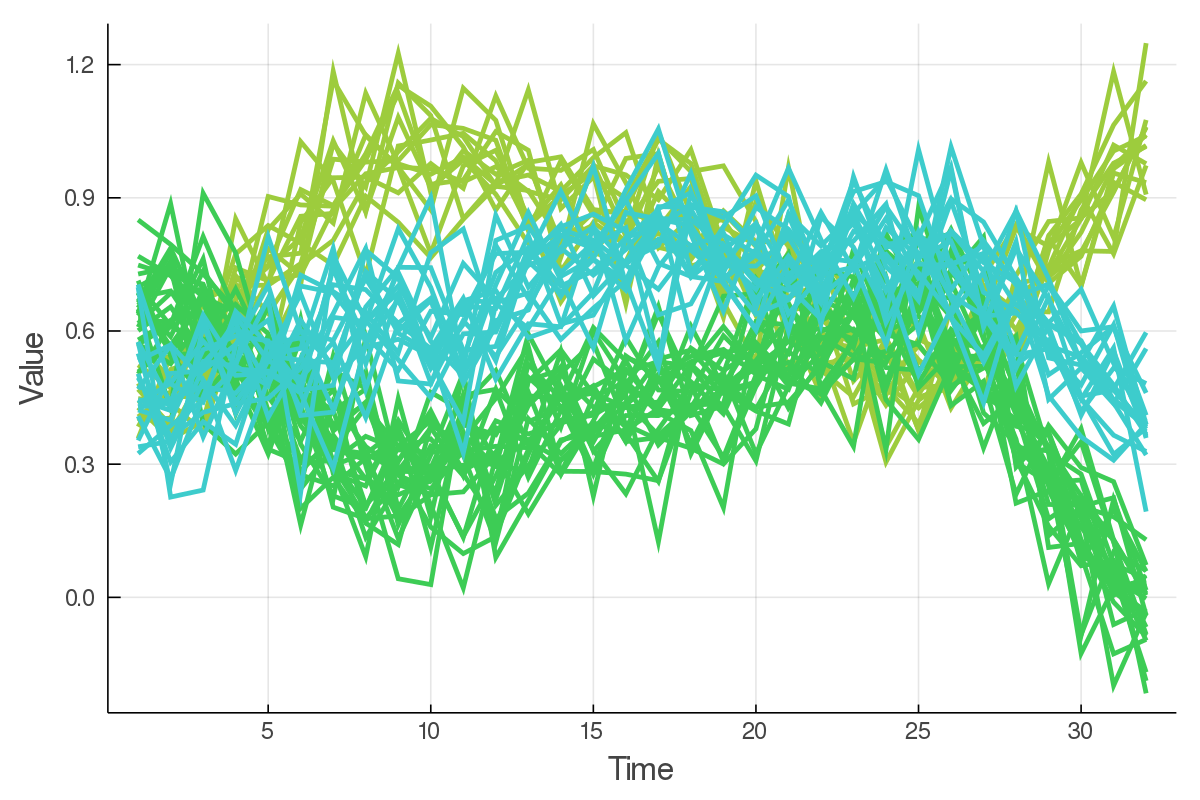}
\includegraphics[width=0.32\textwidth]{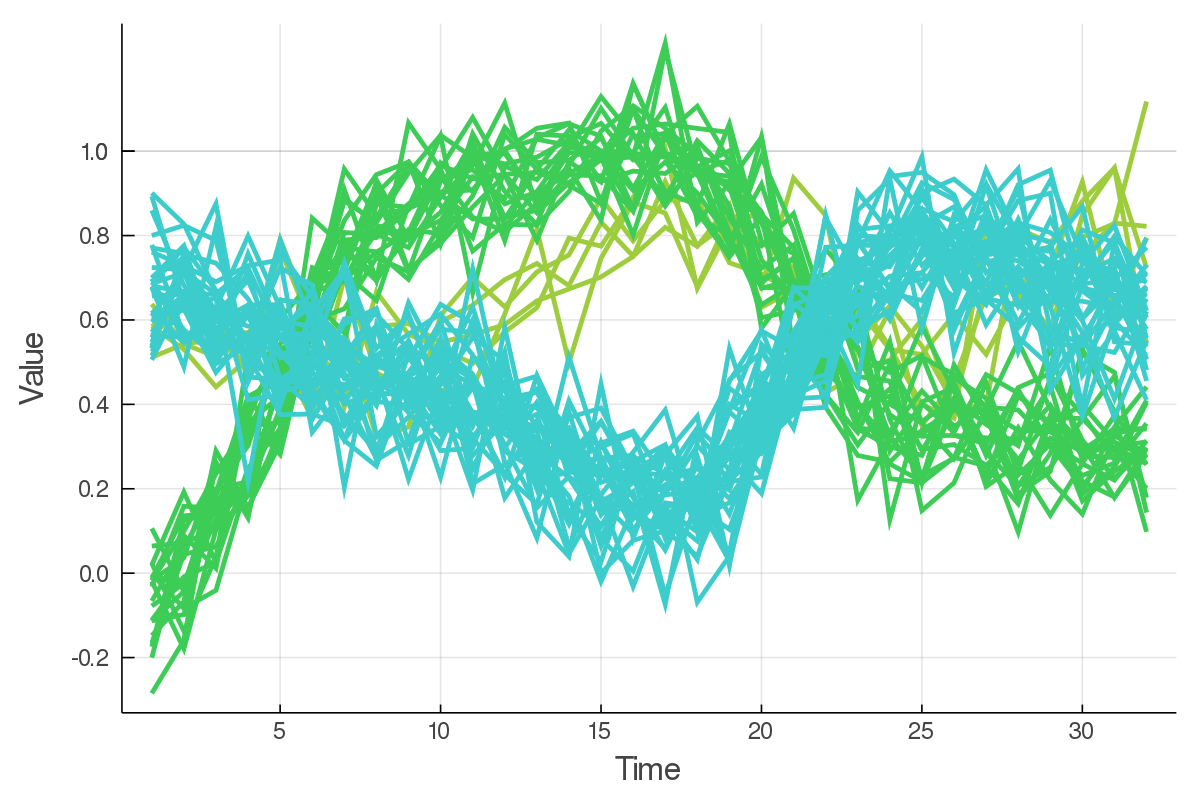}

\includegraphics[width=0.32\textwidth]{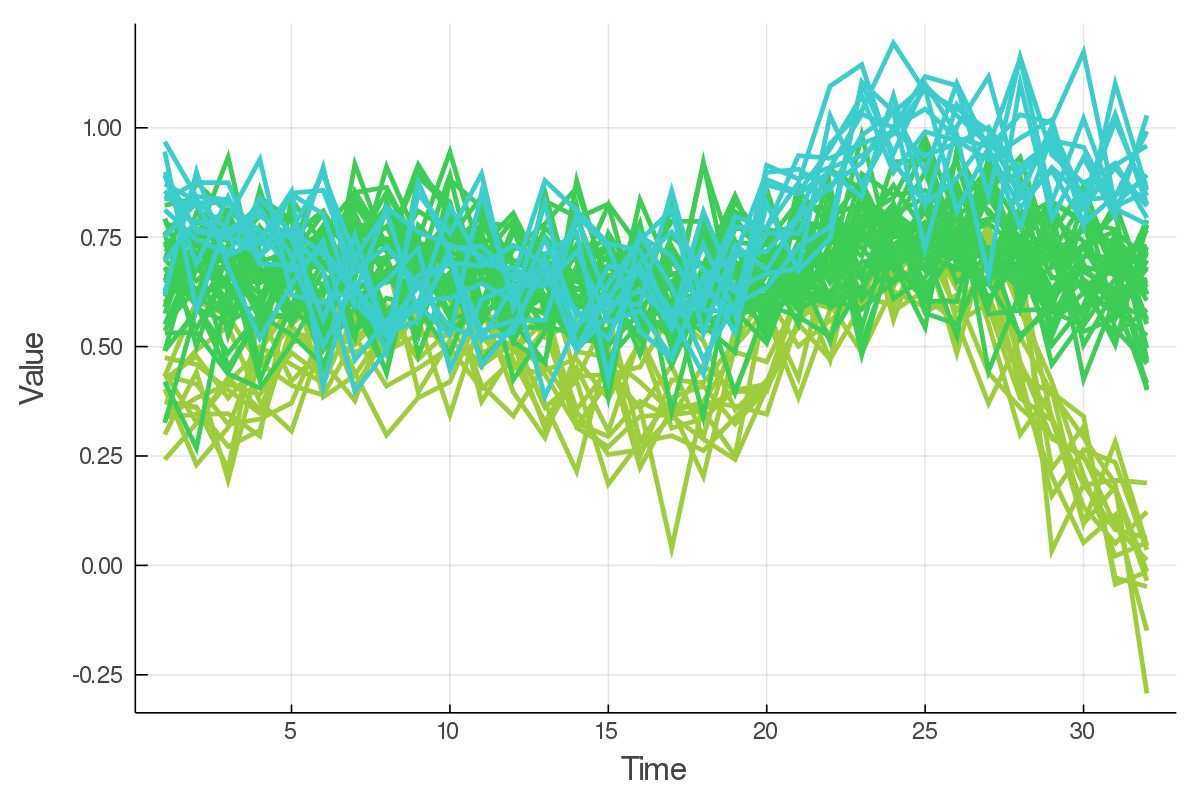}
\includegraphics[width=0.32\textwidth]{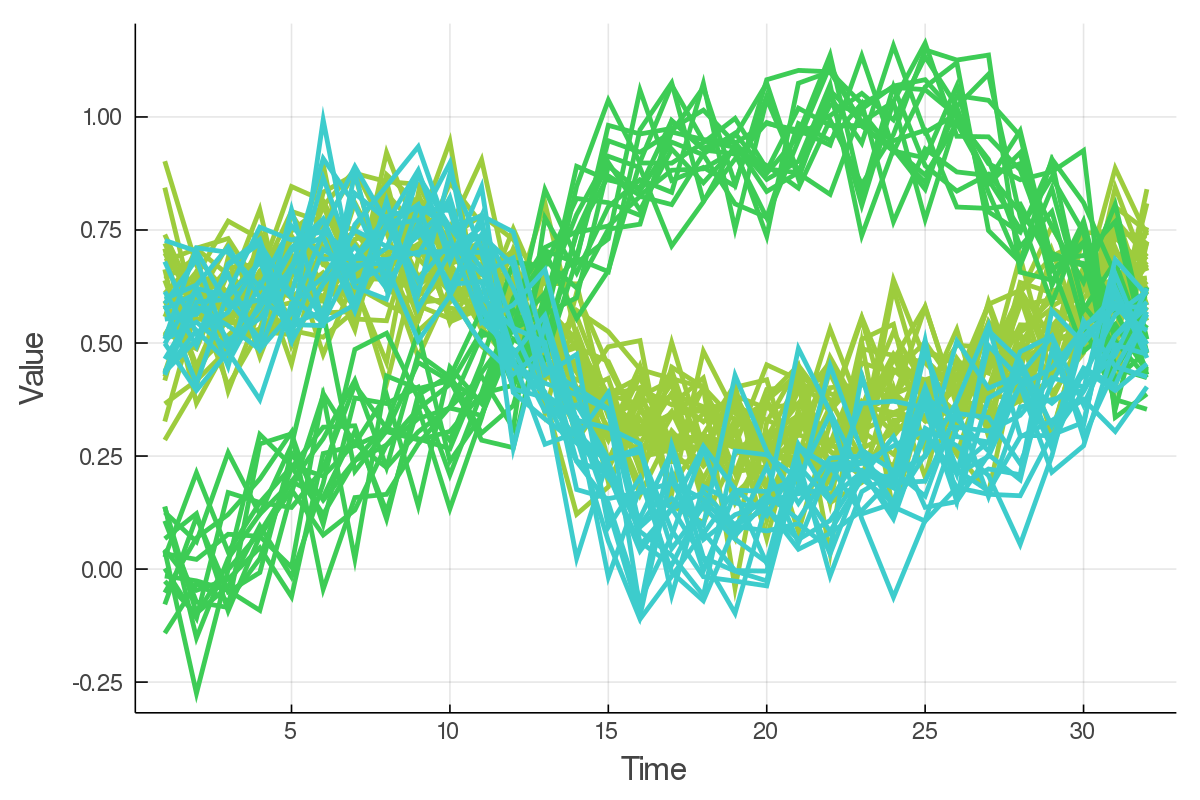}
\includegraphics[width=0.32\textwidth]{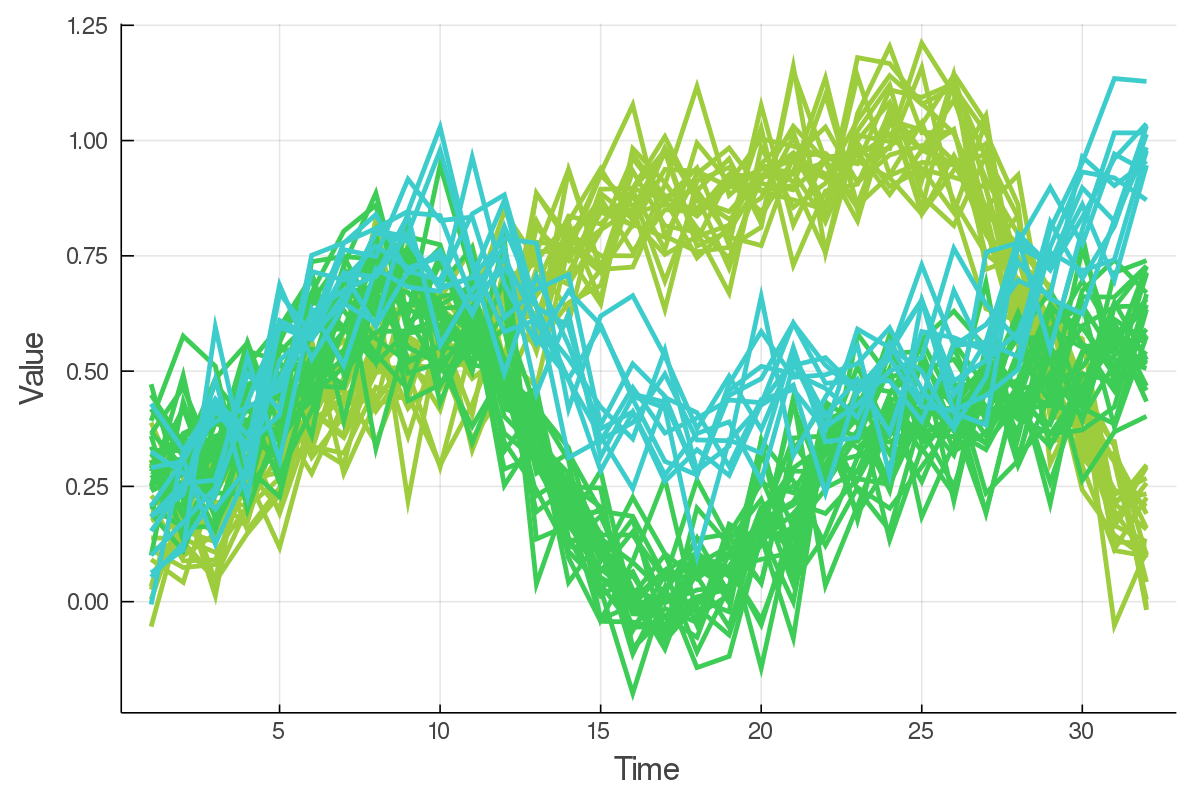}

\includegraphics[width=0.32\textwidth]{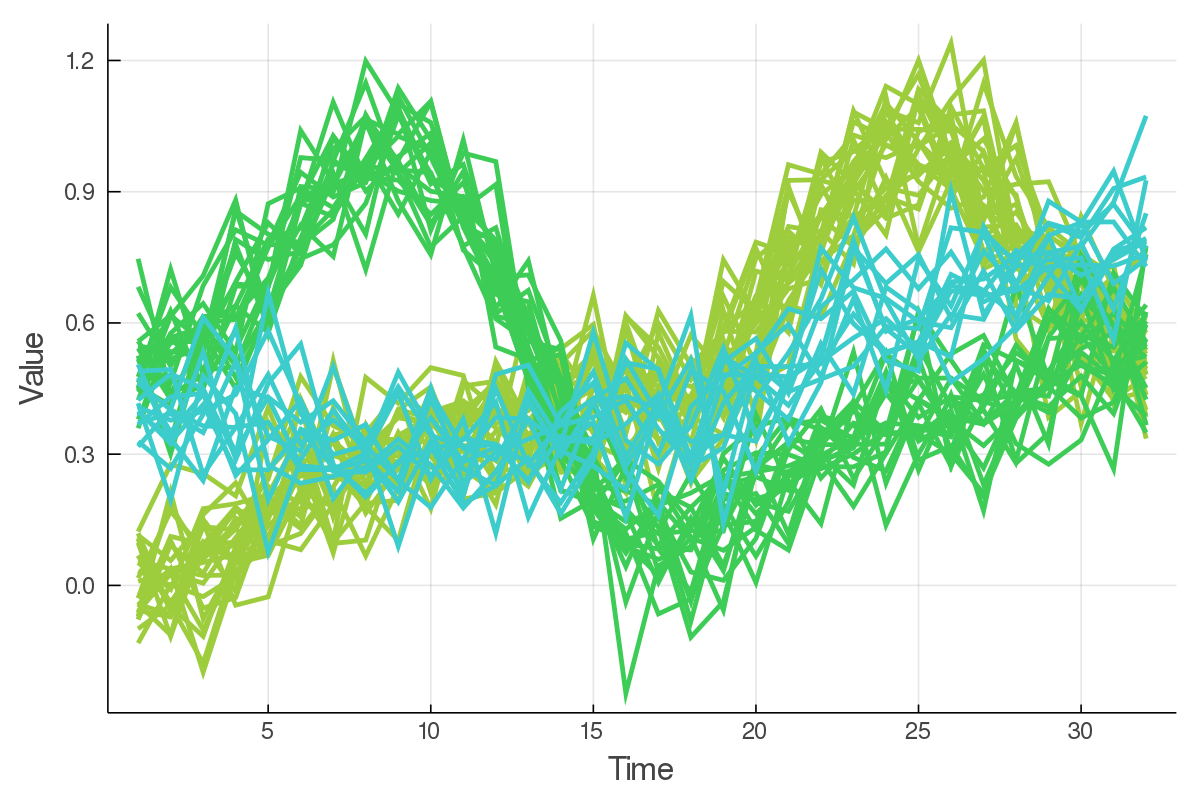}
\includegraphics[width=0.32\textwidth]{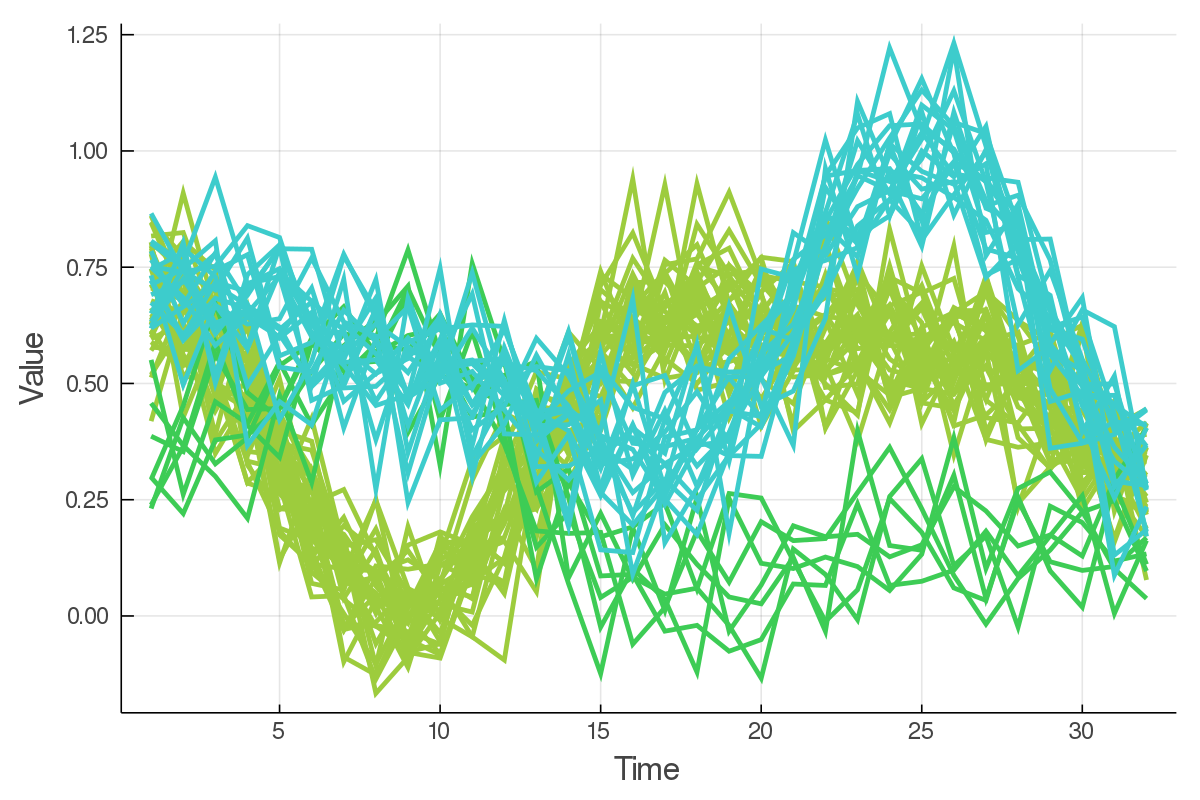}
\includegraphics[width=0.32\textwidth]{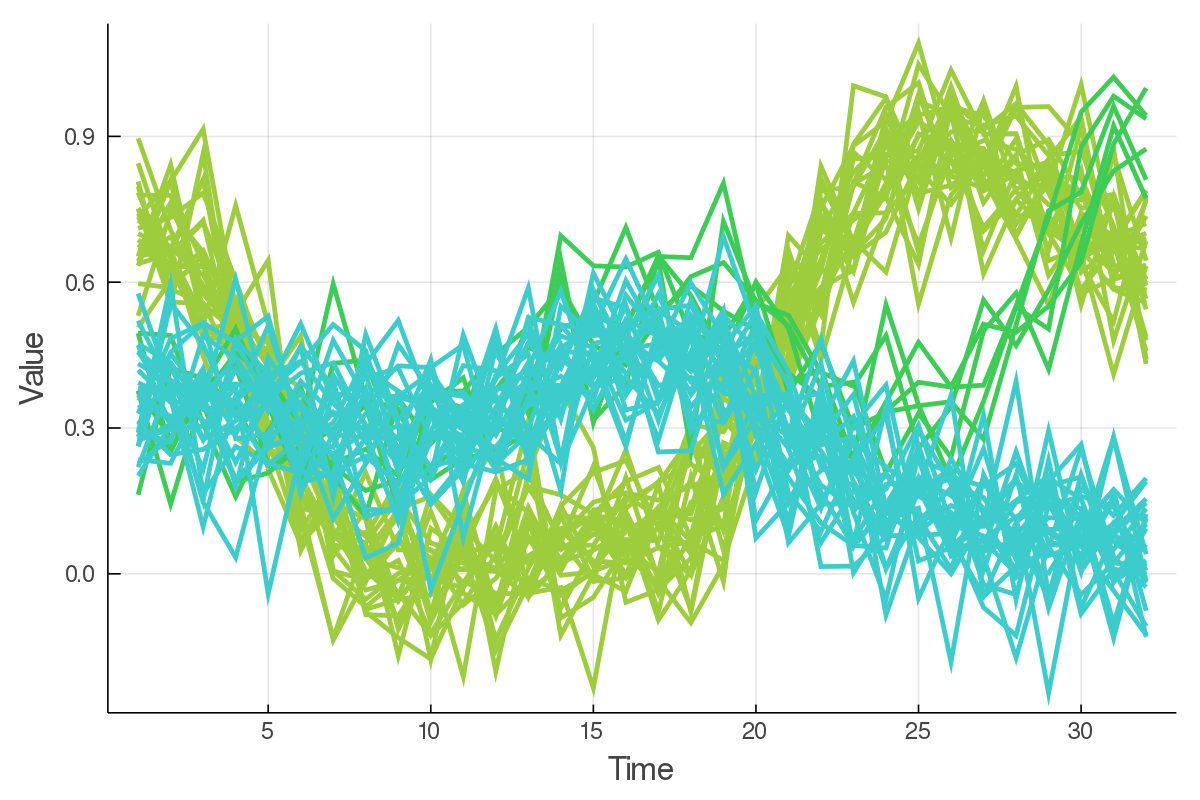}
\end{center}
\caption{Examples of nine randomly generated datasets containing three clusters with a low level of white noise \cite{mlakar2021application}.}
\label{fig:synth}
\end{figure}

For the quantitative analysis we, constructed a synthetic dataset generator which is available on GitHub \cite{smixsgithub}.
The generator constructs a dataset with $c$ clusters, $p$ measurements, and $n$ samples or subjects.
The data has one independent variable time and one dependent variable, measurements.
By adding white noise to their corresponding cluster means, we sample individual subjects from their clusters.
The means are smooth functions created by sampling Perlin noise \cite{perlin1985image}.
We added different levels of noise to simulate low to high distortion in measurements.
Examples of randomly generated datasets can be seen in Figure \ref{fig:synth}. 

\begin{figure}[htb]
    \centering
    \includegraphics[width=\textwidth]{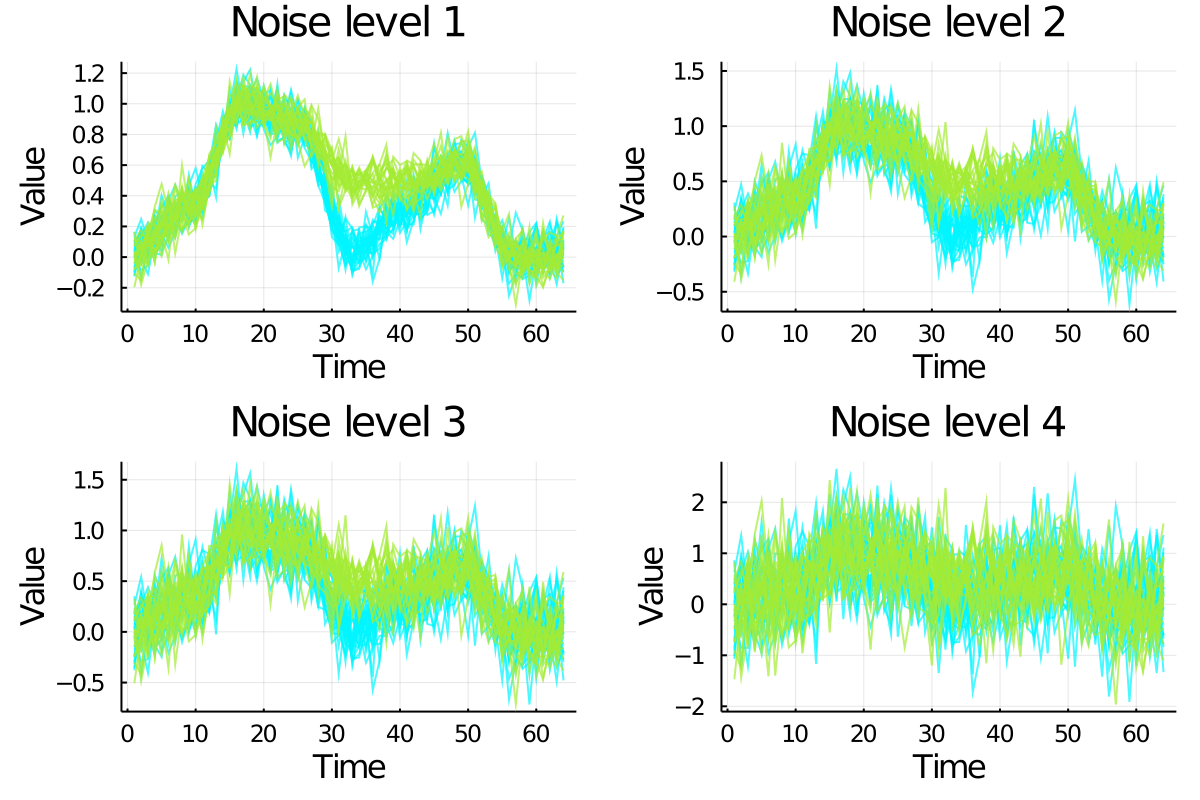}
    \caption{Example of a dataset subjected to increasing amounts of white noise.
    Each color represent samples from a specific latent cluster \cite{mlakar2021application}.}
    \label{fig:synth_noise}
\end{figure}

To test the resilience of the algorithm to noise we vary the amount of noise present in a dataset.
An example of varying noise levels can be seen in Figure \ref{fig:synth_noise}.
We created $500$ different datasets per five different numbers of clusters.
Each of these $2500$ datasets was then subjected to four increasing levels of noise, resulting in $10000$ total datasets.
We used these datasets to conduct the analysis of clustering and regression performance of both SMIXS and GMM.

\subsection{Performance metrics} \label{sec:perf-met}

For the evaluation of clustering performance we use the F-score \cite{chinchor1993muc}, which is defined as
\begin{align*}
    F = 2\frac{P R}{P + R},
\end{align*} 
where $P$ denotes precision and $R$ denotes recall.
The multi-class confusion matrix enables us to compute the F-score for each cluster separately, treating it as a binary classification problem. 
To get the final F-score we average the per cluster F-scores giving equal weights to each cluster independent of their sizes. 
We compare SMIXS and GMM by counting the number of times one of them outperforms the other in terms of F-score. 
We also track the margin of the better-performing algorithm.

We evaluate the quality of the mean cluster curve generated by GMM and SMIXS by calculating the sum of squared differences between the true cluster curves (which are known for synthetic data) and the cluster curves produced by the clustering algorithms.
For the final evaluation and comparison, we average all squared differences of clusters.

Finally, we compare the computational complexities of GMM and SMIXS.
We also analyze the effects that some implemented speed-ups have on the SMIXS algorithm, in particular the addition of the Reinsch algorithm. 
For this evaluation, we examine three versions of SMIXS relative to the base GMM:
\begin{itemize}[label=$\bullet$]
\item SMIXS; complete algorithm as described in Section~\ref{ch:mth} with $\alpha$ optimization and all the time complexity reductions,
\item SMIXS CA; algorithm SMIXS without $\alpha$ optimization, and
\item SMIXS CA-NR; algorithm SMIXS without $\alpha$ optimization and without Reinsch algorithm.
\end{itemize}

We do not evaluate the computational effects of the \citet{hutchinson1985smoothing} algorithm.
The high computational complexity of the base algorithm renders it intractable on the same dataset scale as the above-mentioned variants.
It is safe to say that looking at the theoretical implications of \citet{hutchinson1985smoothing}, \citet{green1993nonparametric} algorithm, and from our own testing, the computational complexity reduction is significant, especially when combined with the Reinsch algorithm \cite{reinsch1967smoothing}.

\subsection{Synthetic evaluation}
\label{sec:syneval}

\subsubsection{Clustering performance analysis}

\begin{figure}[htb]
    \centering
    \includegraphics[width=0.9\textwidth]{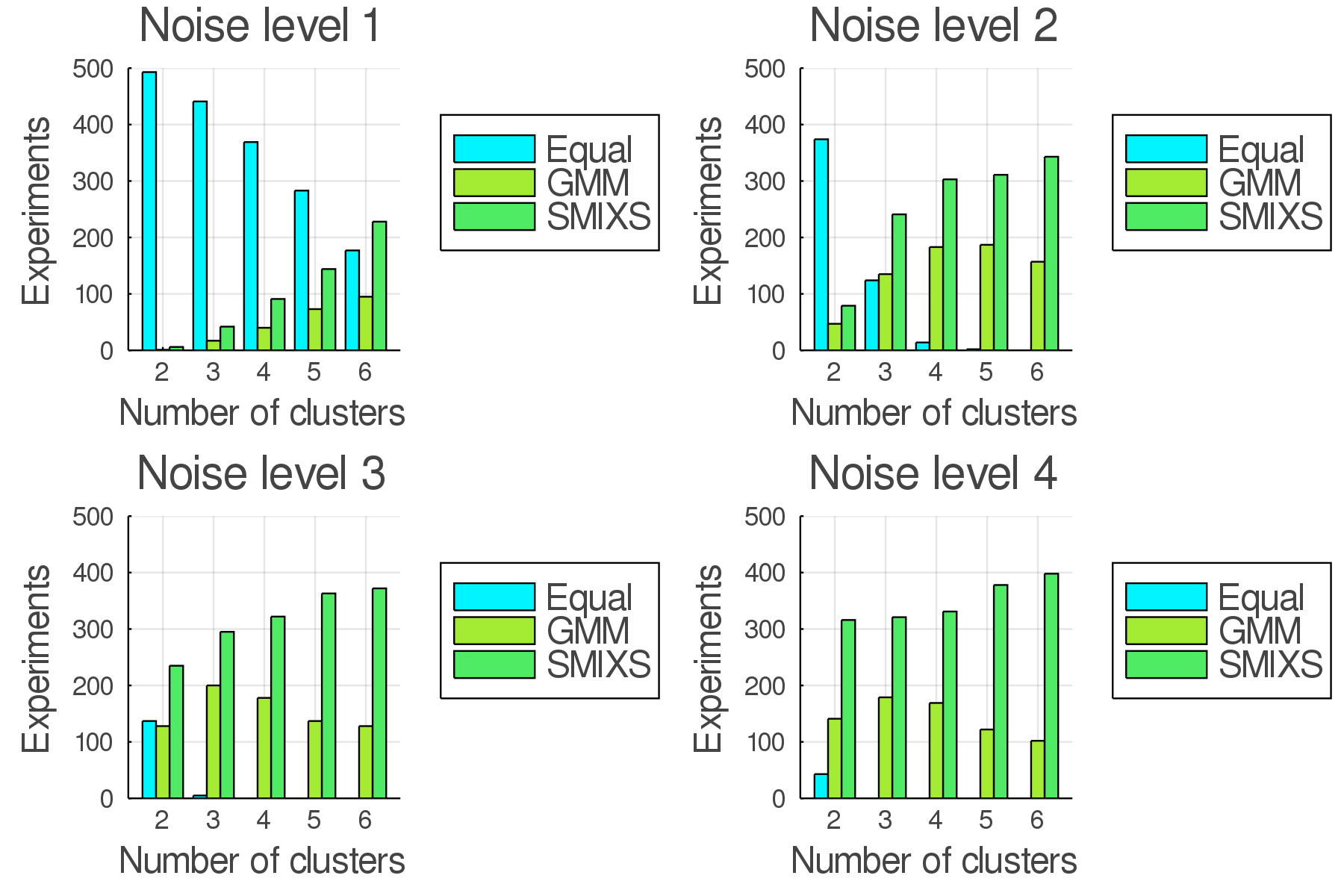}
    \caption{Clustering performance of SMIXS versus GMM. Columns correspond to the amount of datasets where a specific algorithm performed better than the other. The exception is the light blue column which corresponds to the number of datasets where they both performed equally well \cite{mlakar2021application}.}
    \label{fig:synth_cls}
\end{figure}

Looking at Figure \ref{fig:synth_cls} we can see that SMIXS outperforms GMM in almost all scenarios.

\begin{figure}[htb]
\begin{center}
\includegraphics[width=0.33\textwidth]{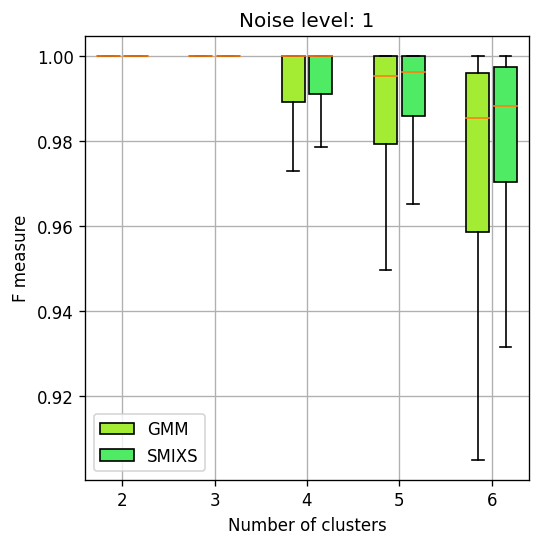}
\includegraphics[width=0.33\textwidth]{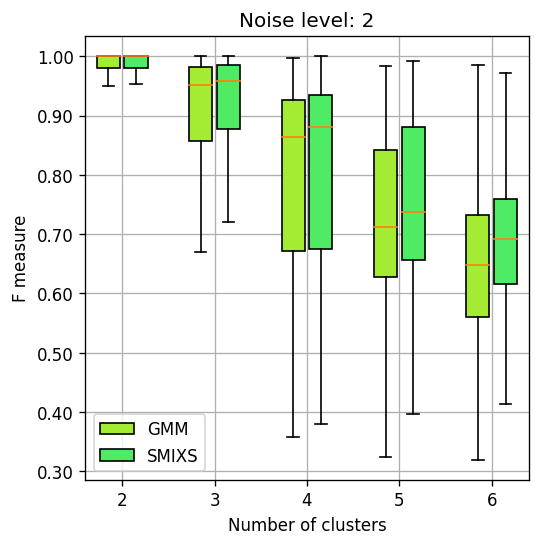}
\includegraphics[width=0.33\textwidth]{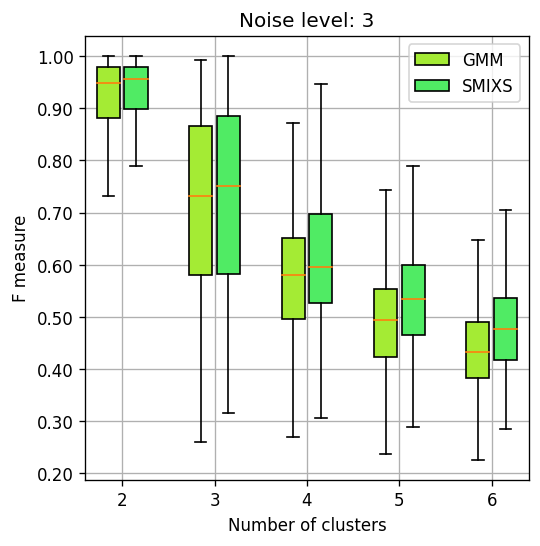}
\includegraphics[width=0.33\textwidth]{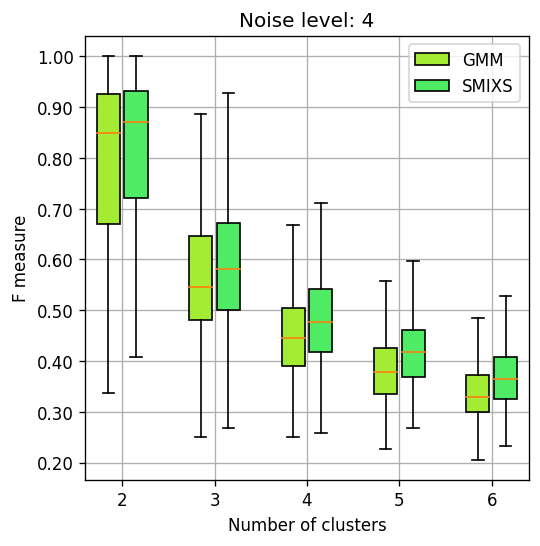}
\end{center}
\caption{Clustering performance of SMIXS and GMM.
The box plots aggregate the results of the respective algorithms over all the datasets for a specific noise level and cluster count.
Higher values denote better performance for that algorithm.}
\label{fig:synth_cls_mrg}
\end{figure}

When the amount of noise is low, which can be seen in noise level one, both algorithms perform equally well for small cluster numbers.
This is due to the clustering problems being too simple such that both algorithms construct a perfect clustering of the dataset.
In those cases, the F-score is the same for both algorithms.
But as we increase the number of clusters in the dataset so does increase the lead of SMIXS over GMM.
By increasing the amount of noise this trend is exacerbated, revealing that SMIXS copes better with a higher number of clusters and high noise situations compared to GMM, while also not performing worse in the case of a smaller number of clusters or low noise.
The margins in performance displayed in Figure \ref{fig:synth_cls_mrg} further corroborate these findings.
The median F-score for SMIXS is constantly above that of GMM.
Likewise, the lower quartiles never extend below those of GMM, and GMM's upper quartiles never extend above SMIXS's.
Again the only exception are the low now noise, low cluster count examples, where both algorithms performed equally well.

\subsubsection{Regression performance analysis}

\begin{figure}[htb]
    \centering
    \includegraphics[width=0.9\textwidth]{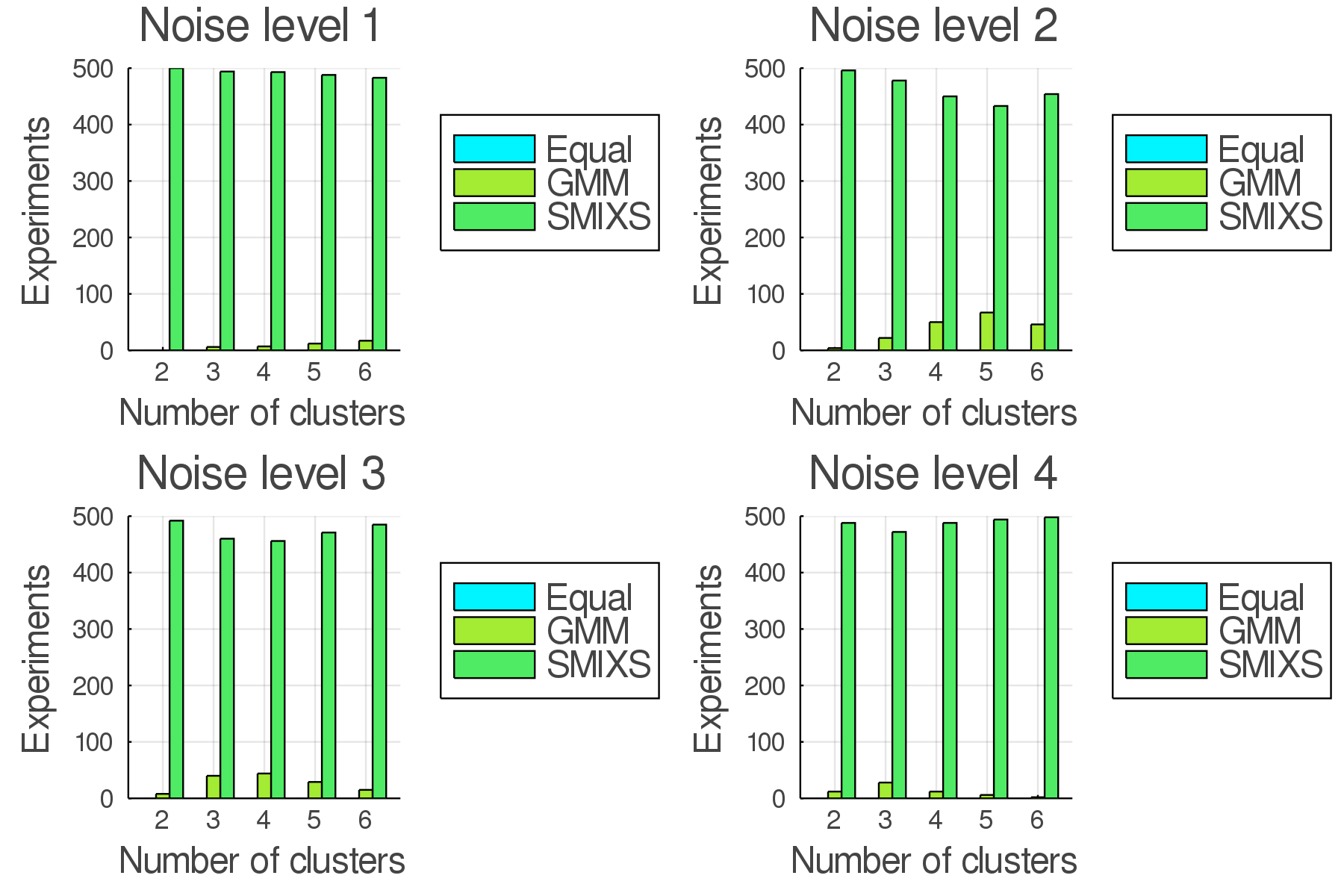}
    \caption{Regression performance of SMIXS versus GMM. Columns correspond to the amount of datasets where a specific algorithm performed better than the other. The exception is the light blue column which corresponds to the number of datasets where they both performed equally well, \cite{mlakar2021application}.}
    \label{fig:synth_rg}
\end{figure}

To investigate the effectiveness of smoothing splines on regression accuracy when the latent generator functions are themselves smooth, we compute the mean squared error between the estimated cluster means compared to the correct cluster means for both GMM and SMIXS.
This will give us an idea of how far from the ground truth the regressed means are.
First, let us look at bar graphs showing the number of datasets where one algorithm outperformed the other in terms of mean squared error (see Figure \ref{fig:synth_rg}).

\begin{figure}[htb]
\begin{center}
\includegraphics[width=0.45\textwidth]{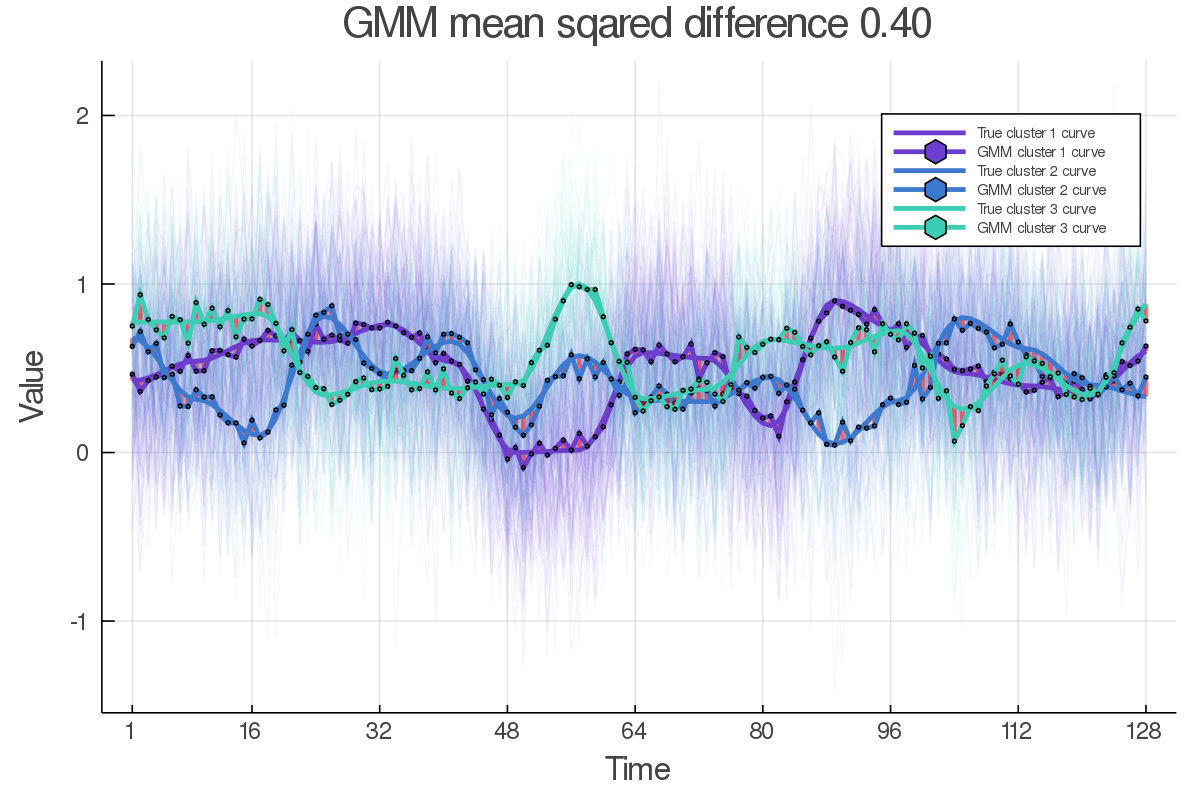}
\includegraphics[width=0.45\textwidth]{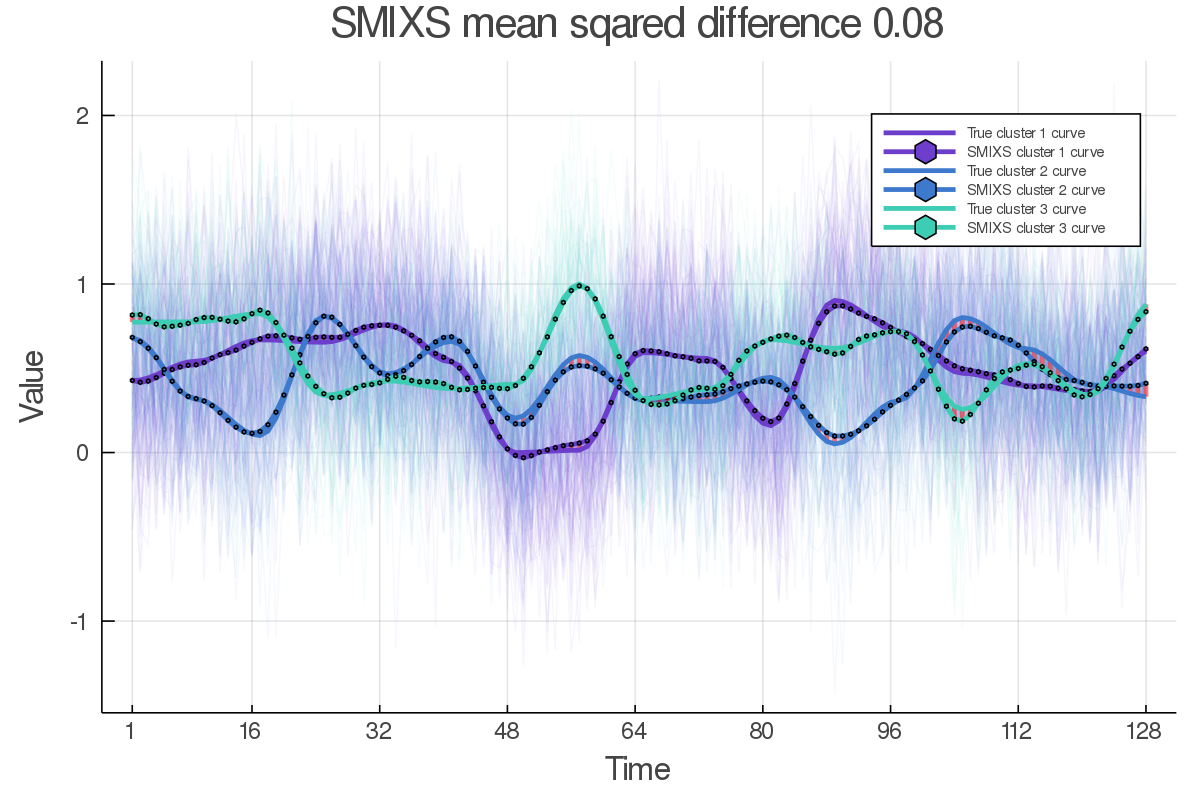}
\end{center}
\caption{Examples of the regression accuracy difference between GMM and SMIXS. SMIXS successfully dampens the effect small noise perturbations have on the regular mean. These perturbations are visible in the regression curve of GMM, \citet{mlakar2021application}.}
\label{fig:synth_rg_ex}
\end{figure}

The difference is more pronounced here compared to the previous analysis.
Here even at low noise levels, and low cluster numbers SMIXS outperforms GMM in a very large number of datasets.
This clearly demonstrates that smoothing splines offer beneficial additions for regressing smooth latent generator functions.
Examples of the regression curves constructed from both GMM and SMIXS, and their comparison to the ground truth are visible in Figure \ref{fig:synth_rg_ex}.


\begin{figure}[htb]
\begin{center}
\includegraphics[width=0.33\textwidth]{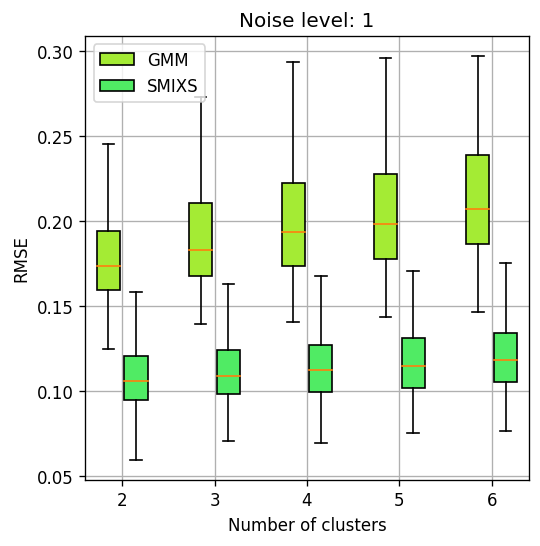}
\includegraphics[width=0.33\textwidth]{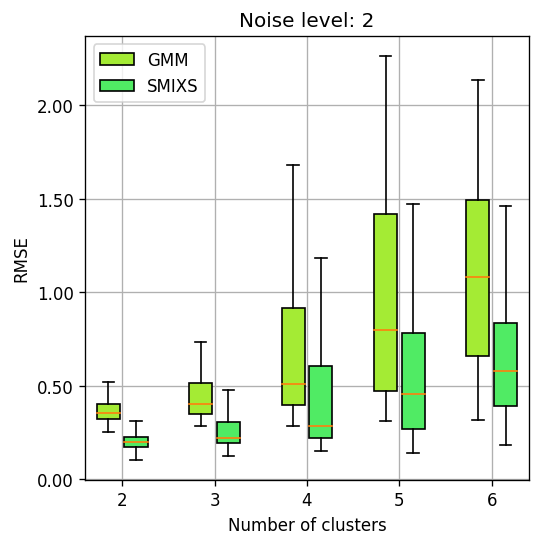}
\includegraphics[width=0.33\textwidth]{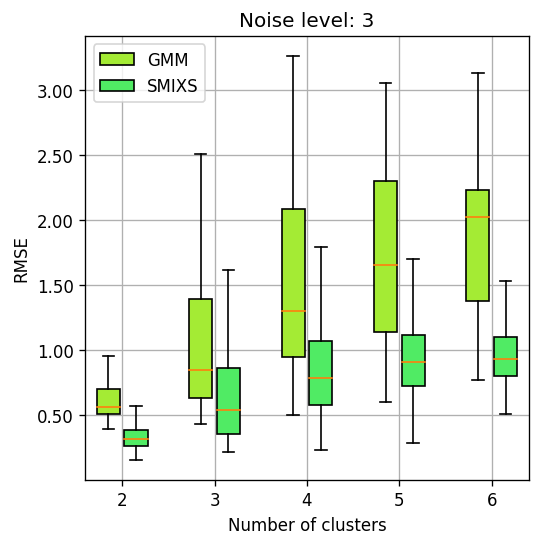}
\includegraphics[width=0.33\textwidth]{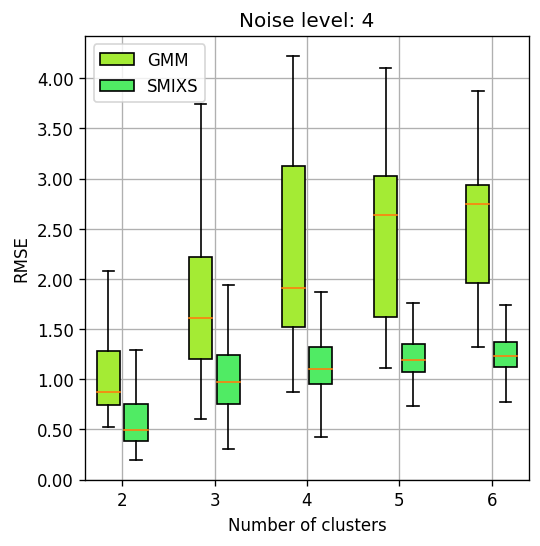}
\end{center}
\caption{Regression performance of SMIXS and GMM. The box plots contain the root mean squared errors (RMSE) between the algorithms computed means and the true cluster curves over all datasets.}
\label{fig:synth_rg_mrg}
\end{figure}

The regression performance in terms of root mean squared error is displayed in Figure \ref{fig:synth_rg_mrg}.
As the amount of noise increases so does the margin in favour of SMIXS, confirming our previous findings that it is the better regression algorithm for smooth latent generating functions. 

\subsubsection{Computational complexity analysis}

\begin{figure}[htb]
\begin{center}
\includegraphics[width=0.45\textwidth]{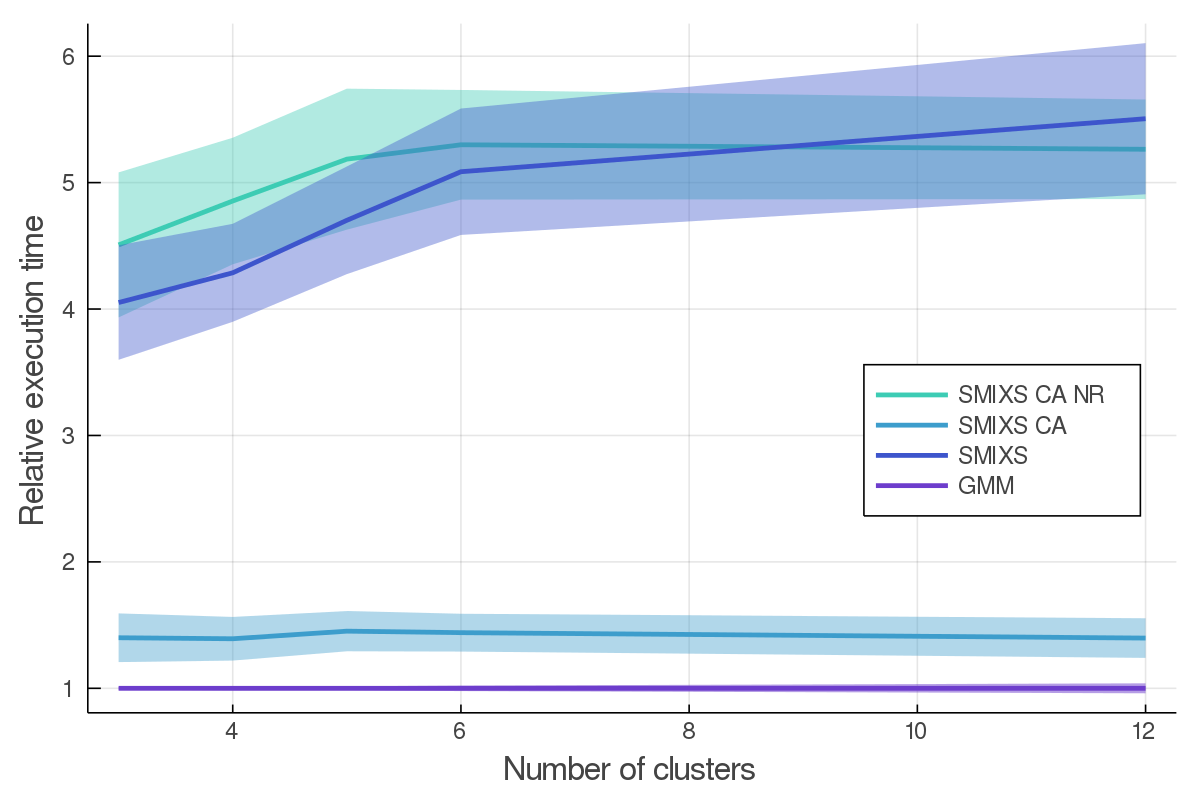}
\includegraphics[width=0.45\textwidth]{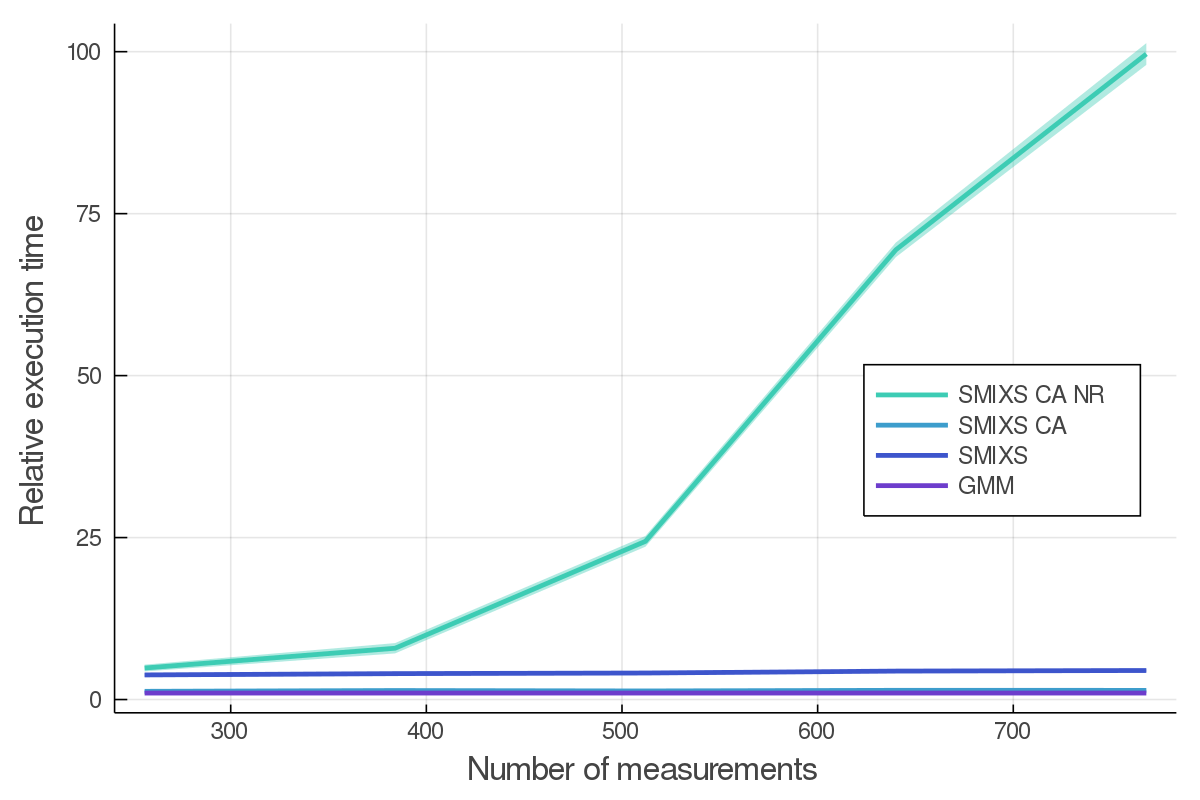}
\includegraphics[width=0.45\textwidth]{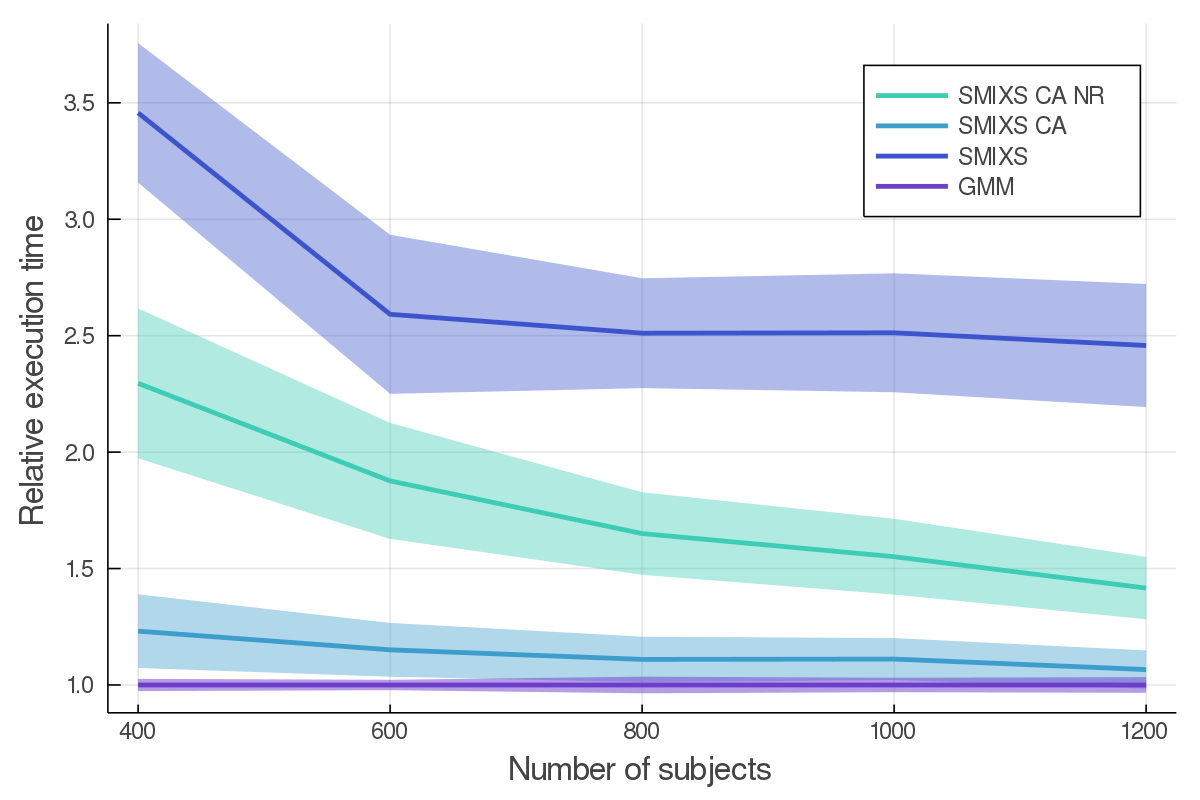}
\end{center}
\caption{Computational complexity of SMIXS variants plotted against GMM.
Each figure represents the relative execution time compared to GMM by varying one parameter in the synthetic dataset.
This allows us to inspect the different impacts dataset parameters have on the algorithm execution, \cite{mlakar2021application}.}
\label{fig:synth_cplx}
\end{figure}

Clustering performance is only a part of the complete performance picture.
We are also interested in how the execution time of the SMIXS algorithm compares to GMM and other variants of SMIXS, namely SMIXS CA-NR, and SMIXS CA (as described in Section \ref{sec:perf-met}).
Therefore, we investigate three scenarios where we create different synthetic datasets with varying numbers of clusters, measurements, and subjects.
This reveals how performance scales with progressively larger datasets.
Observing the results of the analysis displayed in Figure \ref{fig:synth_cplx} we can immediately see that GMM is the fastest of all the tested approaches.
This is not surprising as all the other algorithms constitute an upgrade to GMM.
However, in all tests the SMIXS CA variant follows GMM closely, displaying a relatively small performance hit.
This suggests that if we predetermine a degree of smoothing for a specific dataset, the SMIXS CA algorithm could serve as a valid alternative to GMM even from the execution time perspective (keeping in mind the other performance benefits of smoothing).
Of special note is also the performance of the SMIXS CA-NR variant in the case where we increase the number of measurements. 
At its peak it is almost $100$ times slower than the remaining methods, clearly demonstrating the effectiveness of the Reinsch algorithm in our framework, since SMIXS CA-NR lacks this speed up.

\subsection{Case study-- Clustering of atmospheric sounding data}
\label{ch:cstd}

To show the applicability of SMIXS we use it to cluster atmospheric sounding data from Ljubljana, Slovenia. 
These measurements are an example of longitudinal data, where the air temperature is measured over different pressure levels (altitudes).
Every conducted measurement produces a curve, where the dependent variable corresponds to temperature and altitude represents the independent variable. 
When we record such atmospheric measurements, the effects of different measuring locations, atmospheric states, and measurement times manifest as variations in the data. 
Therefore to extract and encode these variations and potential similarities between individual samples, one can utilize cluster and regression analysis.

Ljubljana is the capital of Slovenia which lies in a basin $295$ meters above sea level with a very unfavorable dispersion situation \cite{pucer2018impact}. 
It exhibits a continental climate with cold winters and hot summers. 
Temperature inversions are common in winter. 
Air temperature usually decreases with the increase of altitude, but when the temperature at the ground is cooler than higher in the atmosphere we can say that we have a temperature inversion. 
With temperature inversion the temperature increases from the ground up to a certain point where it starts decreasing with altitude as expected. 
Temperature inversions affect air quality \cite{pucer2018impact}. 
The main air pollutant measured in Ljubljana is PM$_{10}$ \cite{pucer2018impact}, and despite concentrations decreasing in the last years, on days with temperature inversion and low wind conditions concentrations can still get very high. 
The Slovenian Environment Agency (ARSO) \cite{arso} performs atmospheric sounding \cite{golden1986} every day at 5 in the morning using a radiosonde attached to a weather balloon. 
This way they measure temperature (temperature profile), air humidity, and wind speed at different altitudes, with the mean maximum altitude being $19913.7$ meters above sea level. 
Early morning temperature inversions can easily be identified and characterized by visually inspecting such temperature measurements, but their automatic processing is not straightforward. 
In this case study, we show how SMIXS can be used for the automatic processing of meteorological sounding data by clustering them in several clusters. 
Clustering helps us show that the depth of the morning temperature inversion is associated with higher daily PM$_{10}$ concentrations in Ljubljana. 

The used data consist of $100$ temperature measurements from altitudes of $300$ to $750$ meters above sea level at Ljubljana, capturing the most relevant air layers. 
Temperature inversions usually occur at lower altitudes.
The closer they are to the ground, the more they affect air pollutant concentrations. 
PM$_{10}$ daily concentrations are being measured at the same location as the starting point for atmospheric sounding. The data was provided by ARSO. The data is from the years $2017$ to $2019$, resulting in $1072$ samples.

We assessed the adequate number of cluster with the Bayesian information criterion (BIC) \cite{schwarz1978estimating}.
We observe the plotted BIC curve for an increasing number of clusters from $2$ to $19$ and choose the number of clusters after which the decrease in BIC is not significant enough.
In our case, this was when BIC improved by less than three percent when we increased the number of clusters.
This procedure is in essence a heuristic but it provides a good guideline as to the likely number of optimal clusters, making a trade-off between interpretability (too many clusters are hard to interpret) and cluster homogeneity.
The analysis took $60$ minutes to complete with $50$ initializations per each number of clusters.
Each time we keep the best initialization in terms of BIC .

\begin{figure}[htb]
    \centering
    \includegraphics[width=\textwidth]{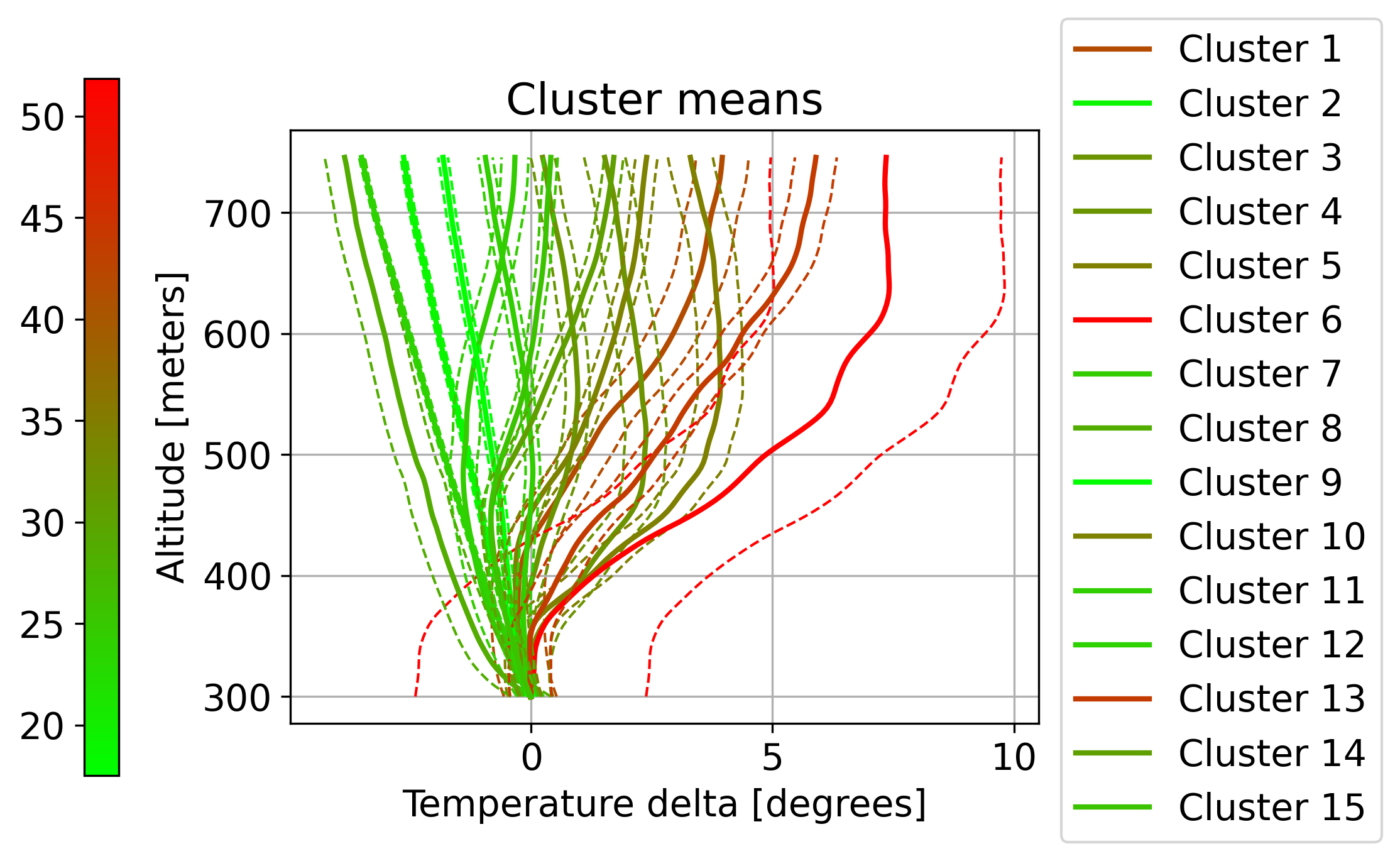}
    \caption{Centroids of the 15 different clusters representing different temperature profiles.
    To produce the $15$ cluster analysis SMIXS required five minutes, conducting $50$ different clusterings and choosing the best one as the final result.}
    \label{fig:centroidi}
\end{figure}

The centroids with the associated standard deviations of the clustered atmospheric sounding temperature profiles are shown in Figure \ref{fig:centroidi}. 
The colour of the centroids represents the measured daily PM$_{10}$ concentration. 
The centroid on the right represents the days with the most extreme temperature inversions and it also represents the most polluted (the reddest) days. 
When we observe the centroids from right to left we can see that clusters on the left are much less polluted (green) than the ones on the right, which is expected. 
When there is an extreme temperature inversion in the morning it is quite usual that it does not break down during the day and air with smog remains trapped near the ground. 
When there is no temperature inversion or a shallow one in the morning the air masses usually mix at some time during the day due to atmospheric convection and then pollution dissipates. 

\begin{figure}[htb]
\begin{tabular}{cc}
 \includegraphics[width=0.412\textwidth]{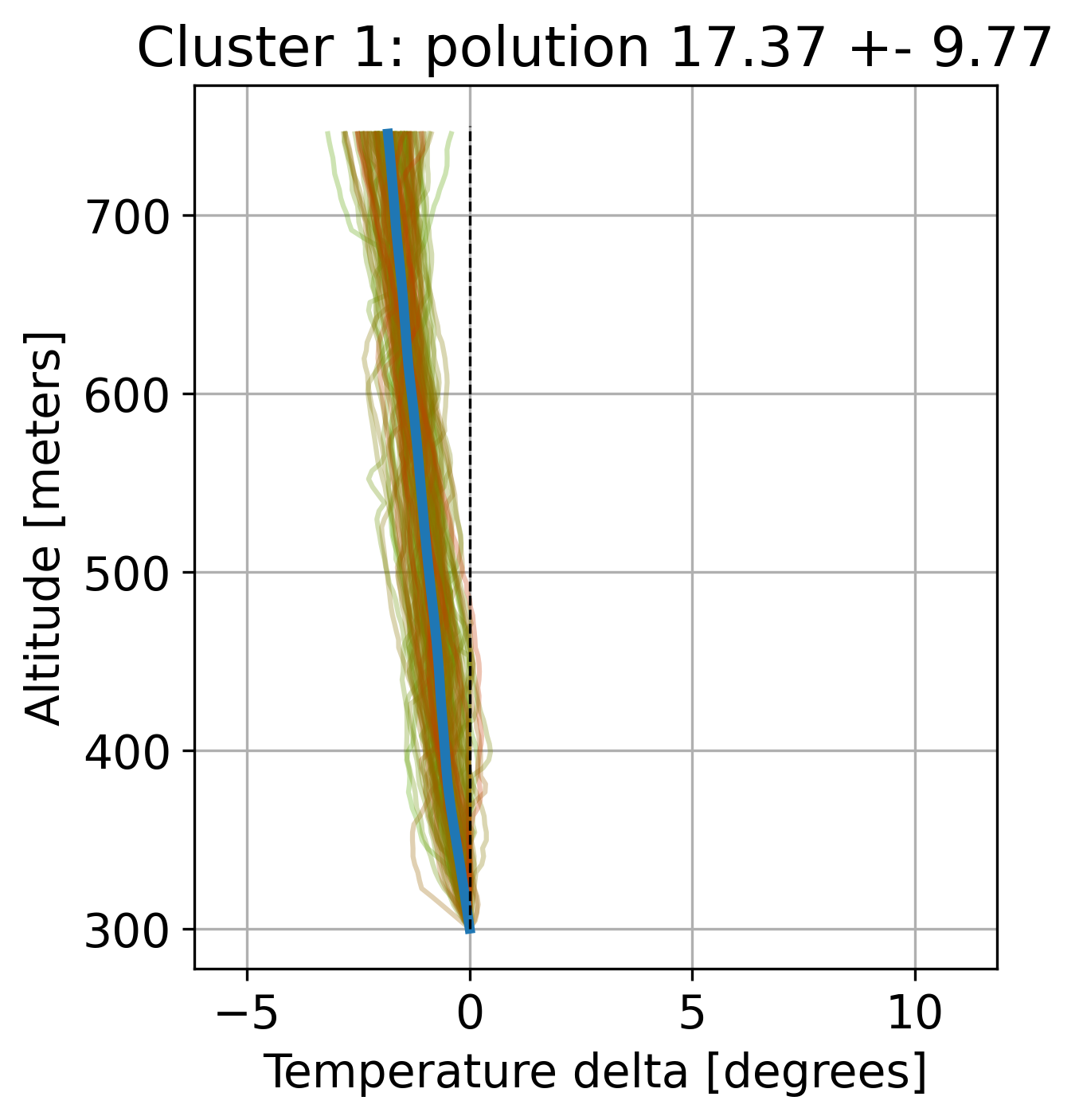} & \includegraphics[width=0.412\textwidth]{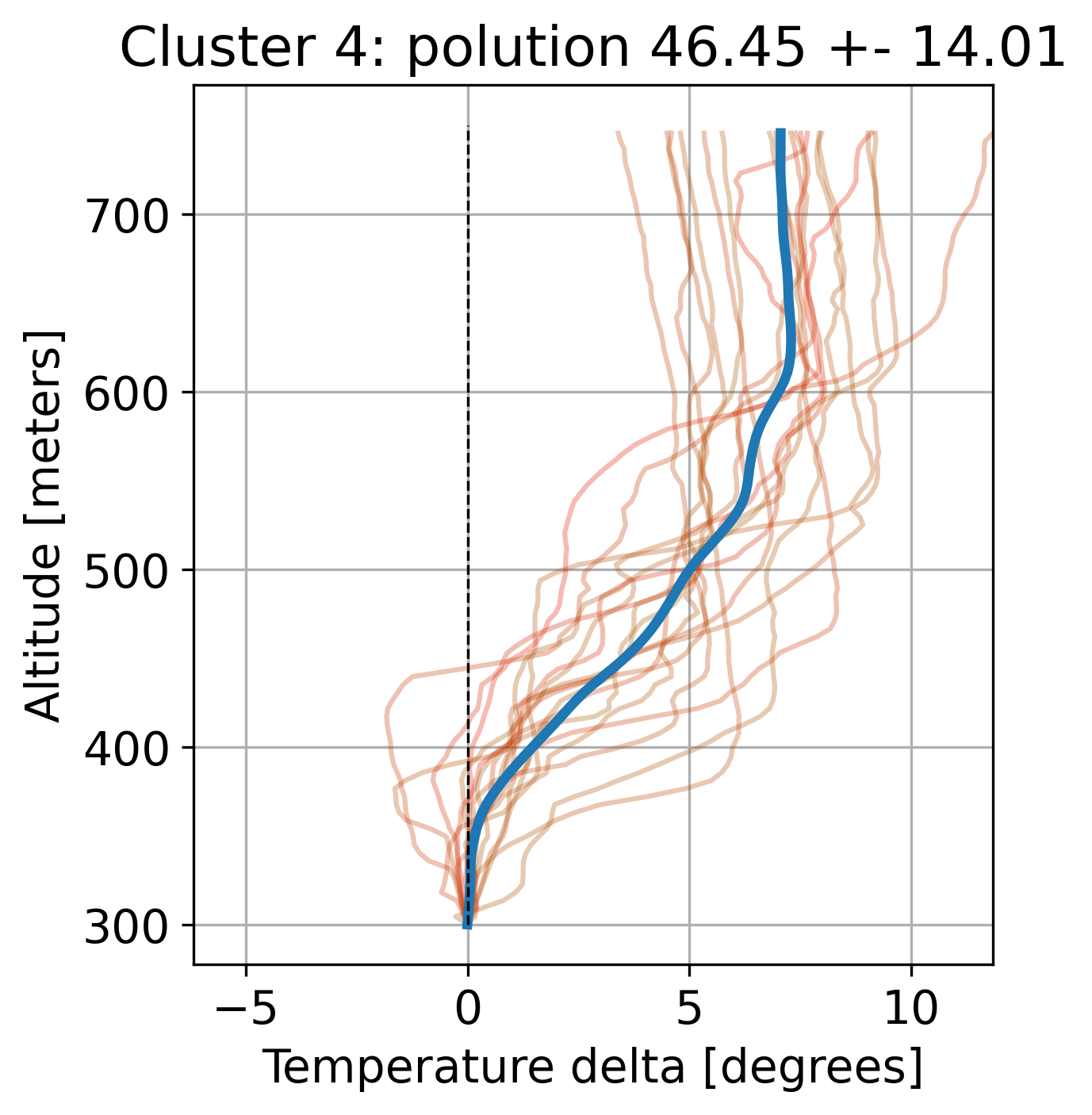}\\
 \includegraphics[width=0.412\textwidth]{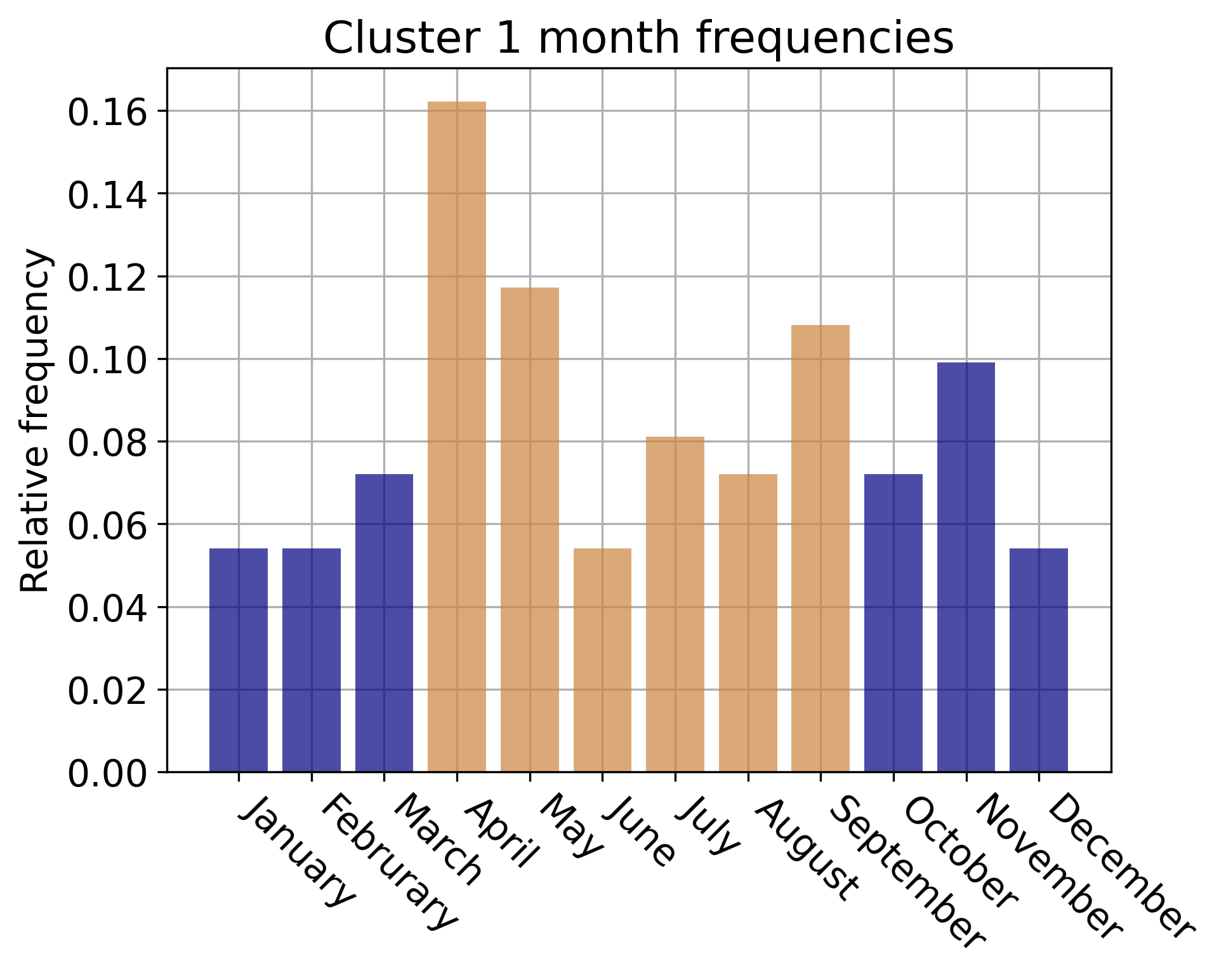} & \includegraphics[width=0.412\textwidth]{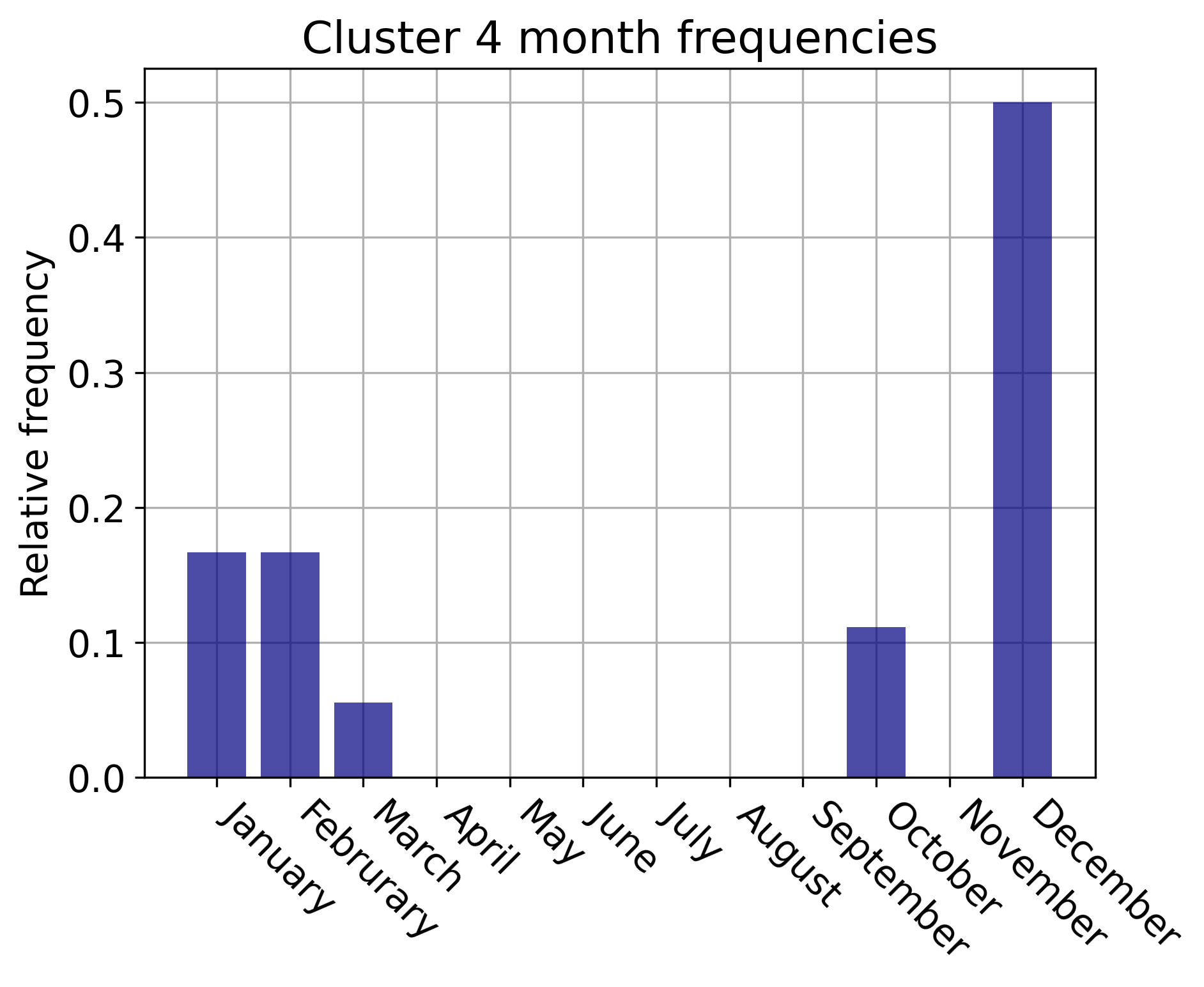}\\
\end{tabular}
\caption{Temperature  profiles from Clusters 1 and 4 as clustered by the SMIXS algorithm (up) and the monthly frequencies of the two clusters (below). Cluster 1 has much more members than Cluster 4.}
 \label{fig:cls1_cls4}
\end{figure}

Figure \ref{fig:cls1_cls4} represents the most and the least polluted clusters. 
Above each plot are the mean cluster PM$_{10}$ concentrations with their associated standard deviation. 
Cluster 1 represents a normal situation where the temperature decreases with increasing altitude, also the PM$_{10}$ concentrations associated with it are predominantly low. 
Cluster 4 comprises extreme temperature inversion profiles, the associated concentrations are high. 
From the sizes of both clusters, we can conclude that the extreme inversion situations are quite rare and that they are typical for the winter months while days with no temperature inversion and low concentrations are common throughout the year, but are more common still in late spring, summer and early autumn.

\begin{figure}[htb]
    \centering
    \includegraphics[width=0.7\textwidth]{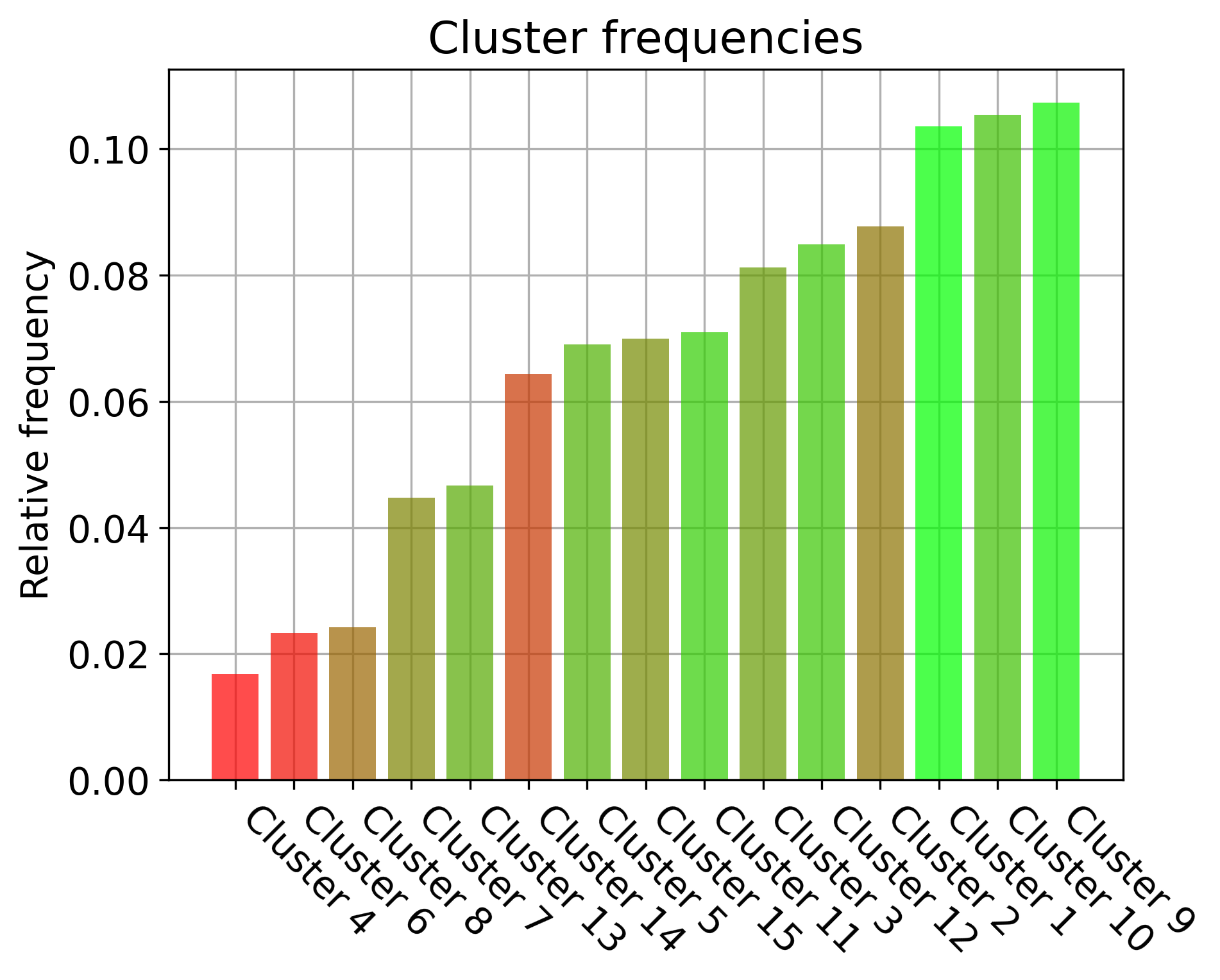}
    \caption{Relative frequencies of the temperature profiles clustered in different clusters. The colour represents the average daily concentration PM$_{10}$}
    \label{fig:frequencies}
\end{figure}

Figure \ref{fig:frequencies} represents the relative frequencies of the temperature profiles in each cluster.
It shows the same as we have seen with Clusters 1 and 4. 
The extreme temperature inversions are quite uncommon, but temperature inversion in general are not (see Clusters 14 and 12). 
Still, days without temperature inversions are much more common especially outside the winter months.

The application of the SMIXS enabled us to characterize different temperature profile types typical for Ljubljana, to assess the frequency of temperature inversions, and the frequencies of different profiles per each month. 
It also enabled us to link different temperature profiles with PM$_{10}$ concentrations. 
All this would be impossible by using raw atmospheric sounding data. 
From this case study, we can see that clustering longitudinal data with SMIXS can help us with their interpretation. 
For future work we could add the cluster information of the meteorological sounding data to the PM$_{10}$ prediction models \cite{faganeli2018bayesian}.

\section{Conclusion}
\label{ch:cncl}

In this work, we proposed improvements to the longitudinal data analysis algorithm originally introduced by \citet{nummi2018semiparametric}.
We implemented computational speed-ups pertaining to smoothing splines and modified them to fit this novel context where they are combined with Gaussian mixture models (GMM).
We also provided a new numerically stable variance estimator which is derived from the Expectation maximization framework.
Lastly, we defined and implemented a new smoothing parameter estimation technique, which enabled our algorithm to exhibit a lower computational complexity compared to other prevalent methods such as grid search, while still yielding good smoothing results.
The algorithm was made available on GitHub \cite{smixsgithub}.

We tested the SMIXS algorithm on a synthetic dataset, where we compared its performance to GMM in terms of regression accuracy, clustering accuracy, and computational complexity.
We showed that SMIXS achieves better results than GMM in both regression and clustering tests but lags behind GMM in computational complexity.
However, by predetermining the magnitude of smoothing enforced on each mixture mean, the execution time of SMIXS follows that of GMM closely, while still retaining the benefits brought by the introduction of smoothing splines into the GMM framework.

Finally, we conducted a case study by analysing the atmospheric sounding data measured over the city of Ljubljana, the capital of Slovenia. 
SMIXS yielded interpretable clusters from a large number of intractable, uninterpretable trajectories. 
The identified clusters showed good correlations with daily PM$_{10}$ concentrations making them a prospective feature for future PM$_{10}$ prediction models.

For potential future research, different methods of selecting the smoothing parameter could be explored.
Gradient descent is but one possibility which still allows for multiple variations (multiple $\alpha_k$ optimization steps per EM iteration, different types of dynamic learning rates).
Striking a balance between computational complexity and estimation accuracy is difficult, which makes this problem challenging to solve.
Another possible avenue of research would be to compare the performance between GMM and SMIXS, much in the same way as we did in our work but, with a smaller number of measurements in the dataset.
In such cases, it can be expected that the amount of required smoothing would increase, therefore potentially allowing for a deeper analysis of the regression performance.

\section*{Acknowledgement}
The authors would like to acknowledge the Slovenian Environment Agency
(ARSO) who provided PM$_{10}$ concentrations and atmospheric sounding data.

This work was supported by the Slovenian Research Agency (ARRS) research core funding  P2-0209 (Jana Faganeli Pucer) and P1-0222 (Polona Oblak).

\clearpage

%
%
\bibliographystyle{splncs04nat}
\bibliography{bibl}

\begin{thebibliography}{25}
\providecommand{\natexlab}[1]{#1}
\providecommand{\url}[1]{\texttt{#1}}
\providecommand{\urlprefix}{URL }
\expandafter\ifx\csname urlstyle\endcsname\relax
  \providecommand{\doi}[1]{doi:\discretionary{}{}{}#1}\else
  \providecommand{\doi}{doi:\discretionary{}{}{}\begingroup
  \urlstyle{rm}\Url}\fi

\bibitem[{ars(2022)}]{arso}
{S}lovenian {E}nvironment {A}gency. \url{https://www.arso.gov.si/en/} (2022),
  accessed: 2022-09-13

\bibitem[{smi(2022)}]{smixsgithub}
{SMIXS} {G}it{H}ub repository. \url{https://github.com/Kepister/SMIXS} (2022),
  accessed: 2022-09-13

\bibitem[{Bolker(2015)}]{bolker2015linear}
Bolker, B.M.: Linear and generalized linear mixed models. Ecological
  statistics: contemporary theory and application pp. 309--333 (2015)

\bibitem[{Chinchor and Sundheim(1993)}]{chinchor1993muc}
Chinchor, N., Sundheim, B.M.: Muc-5 evaluation metrics. In: Fifth Message
  Understanding Conference (MUC-5): Proceedings of a Conference Held in
  Baltimore, Maryland, August 25-27, 1993 (1993)

\bibitem[{Dempster et~al.(1977)Dempster, Laird, and
  Rubin}]{dempster1977maximum}
Dempster, A.P., Laird, N.M., Rubin, D.B.: Maximum likelihood from incomplete
  data via the em algorithm. Journal of the Royal Statistical Society: Series B
  (Methodological) \textbf{39}(1), 1--22 (1977)

\bibitem[{Everitt(2013)}]{everitt2013finite}
Everitt, B.: Finite mixture distributions. Springer Science \& Business Media
  (2013)

\bibitem[{Faganeli~Pucer et~al.(2018)Faganeli~Pucer, Pir{\v{s}}, and
  {\v{S}}trumbelj}]{faganeli2018bayesian}
Faganeli~Pucer, J., Pir{\v{s}}, G., {\v{S}}trumbelj, E.: A {{B}}ayesian
  approach to forecasting daily air-pollutant levels. Knowledge and Information
  Systems \textbf{57}(3), 635--654 (2018)

\bibitem[{Fitzmaurice et~al.(2012)Fitzmaurice, Laird, and
  Ware}]{fitzmaurice2012applied}
Fitzmaurice, G.M., Laird, N.M., Ware, J.H.: Applied longitudinal analysis, vol.
  998. John Wiley \& Sons (2012)

\bibitem[{Genolini and Falissard(2010)}]{clskml}
Genolini, C., Falissard, B.: Kml: k-means for longitudinal data. Computational
  Statistics \textbf{25}(2), 317--328 (2010)

\bibitem[{Golden et~al.(1986)Golden, Serafin, Lally, and Facundo}]{golden1986}
Golden, J., Serafin, R., Lally, V., Facundo, J.: Atmospheric sounding systems.
  In: Mesoscale Meteorology and Forecasting, pp. 50--70, Springer (1986)

\bibitem[{Green and Silverman(1993)}]{green1993nonparametric}
Green, P.J., Silverman, B.W.: Nonparametric regression and generalized linear
  models: a roughness penalty approach. Crc Press (1993)

\bibitem[{Hutchinson and de~Hoog(1985)}]{hutchinson1985smoothing}
Hutchinson, M.F., de~Hoog, F.R.: Smoothing noisy data with spline functions.
  Numerische Mathematik \textbf{47}(1), 99--106 (1985)

\bibitem[{Kom{\'a}rek and Kom{\'a}rkov{\'a}(2013)}]{clskom}
Kom{\'a}rek, A., Kom{\'a}rkov{\'a}, L.: Clustering for multivariate continuous
  and discrete longitudinal data. The Annals of Applied Statistics
  \textbf{7}(1), 177--200 (2013)

\bibitem[{Lu and Lou(2022)}]{clslu}
Lu, Z., Lou, W.: Bayesian consensus clustering for multivariate longitudinal
  data. Statistics in Medicine \textbf{41}(1), 108--127 (2022)

\bibitem[{McLachlan et~al.(2019)McLachlan, Lee, and
  Rathnayake}]{mclachlan2019finite}
McLachlan, G.J., Lee, S.X., Rathnayake, S.I.: Finite mixture models. Annual
  review of statistics and its application \textbf{6}, 355--378 (2019)

\bibitem[{Mlakar(2021)}]{mlakar2021application}
Mlakar, P.: Application of mixed model regression in machine learning  (2021)

\bibitem[{Nummi et~al.(2018)Nummi, Salonen, Koskinen, and
  Pan}]{nummi2018semiparametric}
Nummi, T., Salonen, J., Koskinen, L., Pan, J.: A semiparametric mixture
  regression model for longitudinal data. Journal of Statistical Theory and
  Practice \textbf{12}(1), 12--22 (2018)

\bibitem[{Perlin(1985)}]{perlin1985image}
Perlin, K.: An image synthesizer. ACM Siggraph Computer Graphics
  \textbf{19}(3), 287--296 (1985)

\bibitem[{Pucer~Faganeli and {\v{S}}trumbelj(2018)}]{pucer2018impact}
Pucer~Faganeli, J., {\v{S}}trumbelj, E.: Impact of changes in climate on air
  pollution in {{S}}lovenia between 2002 and 2017. Environmental pollution
  \textbf{242}, 398--406 (2018)

\bibitem[{Reinsch(1967)}]{reinsch1967smoothing}
Reinsch, C.H.: Smoothing by spline functions. Numerische mathematik
  \textbf{10}(3), 177--183 (1967)

\bibitem[{Schwarz(1978)}]{schwarz1978estimating}
Schwarz, G.: Estimating the dimension of a model. The annals of statistics pp.
  461--464 (1978)

\bibitem[{Teuling et~al.(2021)Teuling, Pauws, and van~den
  Heuvel}]{doi:10.1080/03610918.2020.1861464}
Teuling, N.G.P.D., Pauws, S.C., van~den Heuvel, E.R.: A comparison of methods
  for clustering longitudinal data with slowly changing trends. Communications
  in Statistics - Simulation and Computation \textbf{0}(0), 1--28 (2021),
  \doi{10.1080/03610918.2020.1861464},
  \urlprefix\url{https://doi.org/10.1080/03610918.2020.1861464}

\bibitem[{Wang(2014)}]{wang2014generalized}
Wang, M.: Generalized estimating equations in longitudinal data analysis: a
  review and recent developments. Advances in Statistics \textbf{2014} (2014)

\bibitem[{Wu(1983)}]{wu1983convergence}
Wu, C.J.: On the convergence properties of the {{EM}} algorithm. The Annals of
  statistics pp. 95--103 (1983)

\bibitem[{Wu and Zhang(2006)}]{wu2006nonparametric}
Wu, H., Zhang, J.T.: Nonparametric regression methods for longitudinal data
  analysis: mixed-effects modeling approaches. John Wiley \& Sons (2006)

\end{thebibliography}

\section{Appendix}
\label{ch:appendix}

\subsection{On the derivation of the penalty term}

The penalty term stems from the framework of smoothing splines.
First, let us give some intuition as to why we would want to constrain the mixture mean vectors in this way.
Gaussian mixtures for longitudinal data can be thought of as regression models.
More specifically, the mixture mean vectors describe regression trendlines, manifested as cluster means.
However, real-world datasets usually contain noise in their measurements, frequently present in the form of high-frequency variations.
To control the effects noise might have on the data descriptors, we constructed using regression, a smoothness constraint can be applied \cite{green1993nonparametric}.
Consider the mixture mean vectors $\bm \mu_k$ to contain $p$ values, which we sample from a smooth, continuous, twicely differentiable function $\mu_k$ defined on the interval $[t_1,t_p]$.
We wish to quantify how smooth this regression function is, as this relates to the variation or nose in the underlying dataset.
To evaluate this smoothness we can use the following criterion
\begin{equation}
\label{eq:pnlt_smth}
    \int_{t_1}^{t_p} \left (\frac{\partial^2 \mu_k(t)}{\partial t^2} \right)^2 dt.
\end{equation}
It turns out that the smooth function that minimizes this quantity and interpolates a set of points on an interval is the natural cubic spline.
Therefore, we interpret the elements of $\bm \mu_k$ to be values of the natural cubic spline.
The spline knots are constructed at each measurement point $p$.
Using the value-second derivative definition of a natural cubic spline \cite{green1993nonparametric} we can rewrite the smoothness quantifier term (\ref{eq:pnlt_smth}) as 
\begin{align*}
    \int_{t_1}^{t_p} \left (\frac{\partial^2 \mu_k(t)}{\partial t^2} \right)^2 dt = \bm\mu_k^\top\bm G\bm\mu_k.
\end{align*}
Indeed, by using $\bm\mu_k^\top\bm G\bm\mu_k$ as a penalty we punish mixture means with high roughness, due to (\ref{eq:pnlt_smth}).
For more details refer to \cite{green1993nonparametric}.
The added penalty term $\lambda_k$ in (\ref{eq:pnlt}) determines the "exchange rate" between the goodness-of-fit and smoothness of the final spline.
High values of $\lambda_k$ result in smoother mixture means.
Recall that the matrix $\bm G$ is defined as a product of two band matrices \cite{green1993nonparametric} and this is, again, the consequence of using natural cubic splines to conduct the regression.
By using the two band matrices $\bm R$ and $\bm G$ a new relationship can be constructed
\begin{equation}
\label{eq:mu_der}
    \bm Q^\top \bm \mu = \bm R \bm \gamma,
\end{equation}
relating $\bm \mu$ to the vector of the spline's second derivatives $\bm \gamma$ at the knots.
For more details regarding these equalities refer to \cite{green1993nonparametric}.
The choice of using cubic splines is not arbitrary but is a consequence of evaluation smoothness using (\ref{eq:pnlt_smth}).

\subsection{Variance estimator derivation}

The derivation of the variance estimator begins with Equation (\ref{eq:etilde}) where we substitute $\lambda_k$ by $\frac{\alpha_k}{\sigma_k^2}$.
We differentiate the resulting equation with respect to $\sigma_k^2$, resulting in
\begin{align*}
    \frac{\partial \tilde{E}}{\partial (\sigma_k^2)} = \sum_{i = 1}^{n}\hat{z}_{ik} (-p\sigma_k^{-1} + \sigma_k^{-3}(\bm y_i - \bm \mu_k)^\top(\bm y_i - \bm \mu_k)) + \alpha_k\sigma_k^{-3}\bm \mu_k^\top \bm G \bm \mu_k
\end{align*}
and equating the derivative to zero yields
\begin{align*}
    \hat{\sigma}_k^2 = \frac{\sum_{i = 1}^{n}\hat{z}_{ik}(\bm y_i - \bm \mu_k)^{\top}(\bm y_i - \bm \mu_k) + \alpha_k\bm \mu_k^{\top}\bm G\bm \mu_k}{\sum_{i = 1}^{n} \hat{z}_{ik}p}.
\end{align*}

\subsection{Modified Reinsch algorithm}

To introduce the Reinsch algorithm into our framework we begin by rewriting Equation (\ref{eq:mu_est}) as
\begin{align*}
    \left (\bm W_k + \alpha_k \bm G \right) \hat{\bm \mu}_k = \tilde{\bm y}_k,
\end{align*}
By multiplying both sides from the left with $\bm W_k^{-1}$ and decomposing $\bm G$ to the known product, we obtain
\begin{align}\label{eq:gamma-hat}
 \hat{\bm \mu}_k + \alpha_k \bm W_k^{-1} \bm Q \bm R^{-1} \bm Q^\top \hat{\bm \mu}_k = \bm W_k^{-1} \tilde{\bm y}_k.
\end{align}
The matrix $\bm W_k$ is a positive multiple of an identity matrix and is hence invertible. 
In the case where all $\hat{z}_{ik}$ would be zero, we would have an invalid clustering situation where no sample would belong to the mixture $k$.
Excluding such cases, we can proceed by multiplying \eqref{eq:gamma-hat} by $\bm Q^\top$ from the left side and using  \eqref{eq:mu_der} to obtain
 \begin{align*}
     \left ( \bm R + \alpha_k \bm Q^\top \bm W_k^{-1} \bm Q \right ) \hat{\bm \gamma}_k=\bm Q^\top \bm W_k^{-1} \tilde{\bm y}_k.
 \end{align*}
Since matrix $\bm R + \alpha_k \bm Q^\top \bm W_k^{-1} \bm Q$ is symmetric, pentadiagonal, positive-definite matrix, it can be written in its Cholesky decomposition as $\bm L \bm D \bm L^\top$ yielding 
\begin{align*}
    \bm Q^\top \bm W_k^{-1} \tilde{\bm y}_k = \bm L \bm D \bm L^\top \hat{\bm \gamma}_k.
\end{align*}
We can solve this using forward and backward substitution due to $\bm D$ being a band matrix and $\bm L$ being a lower triangular band matrix.

\subsection{Modified Hutchinson and de Hoog algorithm}

In this subsection we will show that
\begin{align*}
    \hat{\bm \mu}_{kj}^{-\{ij\}} -  \bm y_{ij} = \frac{\hat{\bm \mu}_{kj} - \bm y_{ij}}{1 - \bm S_{jj}\hat{z}_{ik}}.
\end{align*}

Our proof will lean on the methodology outlined in \cite{green1993nonparametric}. 
To this end we require a modified vector of samples $\bm y$, namely the vector $\bm y^*_{ab}$ defined as
\begin{align*}
    \bm y^*_{ab} = 
    \begin{cases}
    \hat{\mu}_k^{\{-ab\}}(t_b), & \text{if } a = i, b = j,\\
    \bm y_{ab}, & \text{otherwise.}
    \end{cases}
\end{align*}
First, the value $\hat{\mu}_k$ represents a natural cubic spline defined on the interval $[t_1, t_p]$.
Indeed, we treat the mixture mean vector $\bm \mu_k$ as a vector whose elements are the values of the natural cubic spline $\mu_k$ evaluated at its knots.
Following this, the modified vector of observations is similar to the non-modified one but for one difference.
The $j$-th measurement from the $i$-th sample is replaced by the value of the spline $\hat{\mu}_k^{\{-ij\}}(t_j)$, estimated by removing the $j$-th measurement of the $i$-th sample from the dataset, sampled at the point in time $j$.
Now we can form the following statement
\begin{align}
    & \sum_{a = 1}^{n}\hat{z}_{ak}\sum_{b = 1}^{p}(\bm y^*_{ab} - \mu_k(t_b))^2 + \alpha_k\int_{t_1}^{t_p} \left ( \frac{\partial^2 \mu_k}{\partial t^2} \right)^2 dt \geq \nonumber \\
    & \sum_{a = 1}^{n}\hat{z}_{ak}\sum_{\substack{b = 1 \\ (a,b) \neq (i,j)}}^{p}(\bm y^*_{ab} - \mu_k(t_b))^2 + \alpha_k\int_{t_1}^{t_p} \left ( \frac{\partial^2 \mu_k}{\partial t^2} \right)^2 dt. \label{eq:hoog_0}
\end{align}
Indeed, the two sums are identical but for the omission of a term of the sum from the second line, where the indices $a,b$ equal $i,j$ respectively.
Let us define the minimizer of (\ref{eq:hoog_0}) denoted as $\hat{\mu}^{\{-ij\}}_k$.
This is the minimizer computed by omitting the $j$-th measurement from the $i$-th sample in the dataset.
The following can then be stated
\begin{align}
    & \sum_{a = 1}^{n}\hat{z}_{ak}\sum_{b = 1}^{p}(\bm y^*_{ab} - \mu_k(t_b))^2 + \alpha_k\int_{t_1}^{t_p} \left ( \frac{\partial^2 \mu_k}{\partial t^2} \right)^2 dt \geq \nonumber \\
    & \sum_{a = 1}^{n}\hat{z}_{ak}\sum_{\substack{b = 1 \\ (a,b) \neq (i,j)}}^{p}(\bm y^*_{ab} - \hat{\mu}^{\{-ij\}}_k(t_b))^2 + \alpha_k\int_{t_1}^{t_p} \left ( \frac{\partial^2 \hat{\mu}^{\{-ij\}}_k}{\partial t^2} \right)^2 dt. \nonumber
\end{align}
However, due to the peculiar construction of the vectors $\bm y^*$, the $j$-th measurement of the $i$-th sample is equal to $\hat{\mu}_k^{\{-ij\}}$.
This entails
\begin{align}
    & \sum_{a = 1}^{n}\hat{z}_{ak}\sum_{b = 1}^{p}(\bm y^*_{ab} - \hat{\mu}^{\{-ij\}}_k(t_b))^2 + \alpha_k\int_{t_1}^{t_p} \left ( \frac{\partial^2 \hat{\mu}^{\{-ij\}}_k}{\partial t^2} \right)^2 dt = \nonumber \\
    & \sum_{a = 1}^{n}\hat{z}_{ak}\sum_{\substack{b = 1 \\ (a,b) \neq (i,j)}}^{p}(\bm y^*_{ab} - \hat{\mu}^{\{-ij\}}_k(t_b))^2 + \alpha_k\int_{t_1}^{t_p} \left ( \frac{\partial^2 \hat{\mu}^{\{-ij\}}_k}{\partial t^2} \right)^2 dt. \nonumber
\end{align}
If we now think of the values of $\hat{\mu}^{\{-ij\}}_k$ at the spline knots, we can write those in vector form $\hat{\bm \mu}^{\{-ij\}}_k$, as we are interested in a finite number of measurements.
This vector can be estimated as
\begin{align*}
    \hat{\bm \mu}^{\{-ij\}}_k = \bm S \bm y^*,
\end{align*}
where $\bm S$ is defined as in \eqref{eq:S}.
This is significant as we have derived the relationship between the partial estimator $\hat{\bm \mu}^{\{-ij\}}_k$ and the smoothing matrix $\bm S$. 
Let $\bm S_j$ denote the $j$-th row of the matrix $\bm S$. 
Now let us proceed by deriving the sought after difference (\ref{eq:diff}), starting with
\begin{align*}
    \hat{\bm \mu}^{\{-ij\}}_{kj} - \bm y_{ij} = \bm S_j \sum_{a = 1}^{n}\hat{z}_{ak}\bm y^*_a - \bm y_{ij}.
\end{align*}
Now, separating the $i$-th term from the summation, results in
\begin{align*}
    \hat{\bm \mu}^{\{-ij\}}_{kj} - \bm y_{ij} = \bm S_j \sum_{a = 1 \land a \neq i}^{n}\hat{z}_{ak}\bm y_a + \bm S_j\hat{z}_{ik}\bm y^*_i - \bm y_{ij},
\end{align*}
where the summation, devoid of the the $i$-th sample, transforms to the summation over the non-modified observation vectors $\bm y$.
This is again due to the way $\bm y^*$ are constructed.
By subtracting and adding the term $\bm S_j\hat{z}_{ik}\bm y_i$ to the right side of the above equality we get
\begin{align*}
    \hat{\bm \mu}^{\{-ij\}}_{kj} - \bm y_{ij} = \bm S_j \sum_{a = 1}^{n}\hat{z}_{ak}\bm y_a + \bm S_j\hat{z}_{ik}\bm y^*_i - \bm y_{ij} - \bm S_j\hat{z}_{ik}\bm y_i.
\end{align*}
The summation $\bm S_j \sum_{a = 1}^{n}\hat{z}_{ak}\bm y_a$ is equal to the $j$-th element of the mean vector estimated from the complete, non-modified dataset
\begin{align*}
    \hat{\bm \mu}^{\{-ij\}}_{kj} - \bm y_{ij} = \hat{\bm \mu}_{kj} + \bm S_j\hat{z}_{ik}\bm y^*_i - \bm y_{ij} - \bm S_j\hat{z}_{ik}\bm y_i.
\end{align*}
The dot products $\bm S_j\hat{z}_{ik}\bm y^*_i$ and $\bm S_j\hat{z}_{ik}\bm y_i$ are equal in all but the $j$-th terms resulting in 
\begin{align*}
    \hat{\bm \mu}^{\{-ij\}}_{kj} - \bm y_{ij} = \hat{\bm \mu}_{kj} - \bm y_{ij} + \bm S_j\hat{z}_{ik}(\hat{\mu}_k^{\{-ij\}}(t_j) - \bm y_i).
\end{align*}
Since $\hat{\mu}_k^{\{-ij\}}$ is sampled at the spline knot $t_j$ this essentially equals $\hat{\bm \mu}_{kj}^{\{-ij\}}$, leading us to the final equality
\begin{align*}
    \hat{\bm \mu}_{kj}^{\{-ij\}} -  \bm y_{ij} = \frac{\hat{\bm \mu}_{kj} - \bm y_{ij}}{1 - \bm S_{jj}\hat{z}_{ik}}.
\end{align*}
What remains to be optimized is the computation of the diagonal elements.
Keeping in mind that the matrix $\bm W_k$ is invertible in non-degenerate clustering cases, the calculation of the diagonal elements follows the standard procedure as defined by Hutchinson and de Hoog \cite{hutchinson1985smoothing}.

\end{document}